\documentstyle[preprint,eqsecnum,aps,epsf]{revtex}
\newif\iftightenlines\tightenlinesfalse
\tightenlines\tightenlinestrue

\def\eslt{\not\!\!{E_T}}
\def\to{\rightarrow}

\def\te{\tilde e}

\def\tu{\tilde u}

\def\tb{\tilde b}

\def\tnr{\tilde{\nu}_R}
\def\td{\tilde d}
\def\tQ{\tilde Q}
\def\tL{\tilde L}
\def\tH{\tilde H}
\def\tst{\tilde t}
\def\ttau{\tilde \tau}

\def\tg{\tilde g}
\def\tnu{\tilde\nu}
\def\tell{\tilde\ell}
\def\tq{\tilde q}
\def\tw{\widetilde W}

\def\tz{\widetilde Z}
\begin{document}
\draft
\preprint{\vbox{\baselineskip=14pt%
   \rightline{FSU-HEP-001104}
   \rightline{UH-511-979-01}
   \rightline{MAD/PH-01-1218}
}}
\title{ASPECTS OF SUPERSYMMETRIC MODELS WITH A \\
RADIATIVELY DRIVEN INVERTED MASS HIERARCHY}
\author{Howard Baer$^1$, Csaba Bal\'azs$^2$, Michal Brhlik$^3$, \\
Pedro Mercadante$^1$, Xerxes Tata$^2$ and Yili Wang$^2$}
\address{
$^1$Department of Physics,
Florida State University,
Tallahassee, FL 32306 USA
}
\address{
$^2$Department of Physics and Astronomy,
University of Hawaii,
Honolulu, HI 96822, USA
}
\address{
$^3$Department of Physics,
University of Wisconsin,
Madison, WI 53706, USA
}
\date{\today}
\maketitle
\begin{abstract}

A promising way to reconcile naturalness with a decoupling solution to
the SUSY flavor and CP problems is suggested by models with a
radiatively driven inverted mass hierarchy (RIMH).  The RIMH models
arise naturally within the context of SUSY $SO(10)$ grand unified
theories. In their original form, RIMH models suffer from two problems:
1.) obtaining the radiative breakdown of electroweak symmetry, and 2.)
generating the correct masses for third generation fermions. The first
problem can be solved by the introduction of $SO(10)$ $D$-term
contributions to scalar masses. We show that correct fermion masses can
indeed be obtained, but at the cost of limiting the magnitude of the
hierarchy that can be generated.
We go on to compute predictions for the
neutralino relic density as well as for the rate for the decay $b\to
s\gamma$, and show that these yield significant constraints on model
parameter space.  We show that only a tiny corner of
model parameter space is accessible to Fermilab Tevatron searches,
assuming an integrated luminosity of 25~$fb^{-1}$.
We also quantify the reach of
the CERN LHC collider for this class of models, and find values of
$m_{\tg}\sim 1600$ GeV to be accessible assuming just 10 fb$^{-1}$
of integrated luminosity.
In an Appendix, we list the two loop
renormalization group equations for the MSSM plus right handed neutrino
model that we have used in our analysis.

\end{abstract}

\medskip



\section{Introduction}
Weak scale supersymmetry (SUSY) \cite{review} provides
a highly motivated framework for physics beyond the Standard Model (SM).
The hunt for the predicted
supersymmetric matter is one of the primary tasks for collider
experiments in the 21st century. The details of superparticle signatures
are model-dependent, and may shed light on the mechanism for the communication
of supersymmetry breaking to the observable sector consisting of SM
particles and their superpartners.

An elegant approach is to assume supersymmetry is spontaneously broken
in a ``hidden sector'', and that supersymmetry breaking is communicated
to the observable sector via gravitational interactions\cite{sugra}.
Most generally, in these supergravity models, soft SUSY breaking (SSB)
scalar mass parameters are flavor dependent\cite{sugmas}, and lead to
violation of experimental limits on flavor changing (and possibly also
CP violating) phenomena \cite{masiero}. Generic predictions for CP
violating, but flavor conserving, quantities such as the electric dipole
moments of the electron or the neutron also violate experimental upper
limits on their values. These are the well known SUSY flavor and CP
problems.  The proposal of {\it universality} of scalar masses solves
the SUSY flavor problem, but is not a direct consequence of supergravity
models. Even if universality can be arranged at tree level, quantum
corrections will lift the degeneracy, and lead back to problems in the
flavor and CP sectors of the theory\cite{sugrad}. Problems with flavor
conserving, but CP violating, observables arise because new phases are
possible in SUSY theories. These problems are ameliorated if the
associated (combination of) phases are small, or if sparticles are
sufficiently heavy.

Over the years, considerable effort has been spent on
the search for novel SUSY breaking schemes which are consistent with
FCNC and CP violating processes. Three
prominent classes of models include gauge-mediated SUSY breaking
(GMSB)\cite{gmsb}, anomaly-mediated SUSY breaking (AMSB)\cite{amsb} and
gaugino-mediated SUSY breaking\cite{gomsb}. Each of these models must
introduce additional visible sector fields and/or extra dimensions,
unlike the case with gravity-mediated SUSY breaking models. In these
frameworks, the flavor problem is solved\footnote{Care must be taken to
ensure that the additional fields do not induce new sources of flavor
violation.} because particles with the
same gauge quantum numbers are degenerate, but the CP problem may
remain.

A different solution to the SUSY flavor problem in four dimensional
supergravity theories is {\it decoupling}\cite{decoup}, where the
superpartner masses (for the first two generations) are raised to such
high values that FCNC and CP violating processes are strongly
suppressed. Naively, this solution conflicts with naturalness
considerations\cite{nat}: the SUSY flavor and CP problems demand
multi-TeV scalar masses, while naturalness prefers sub-TeV SUSY breaking
masses.  A proposal which reconciles these requirements is to
arrange for an {\it inverted mass hierarchy} (IMH) for scalars in the
theory: multi-TeV masses for first and second generation scalars, but
sub-TeV masses for third generation and Higgs scalars\cite{imh}.

If the IMH of scalars already exists at the GUT scale (GSIMH models),
then it has been emphasized that two-loop contributions to the
renormalization group evolution of soft SUSY breaking scalar
masses can cause a breakdown in color and/or electric charge symmetry
unless the GUT scale top squark mass is sufficiently large\cite{am}.  Indeed
one can generate viable GSIMH models\cite{gsimh} only over limited
portions of model parameter space.  The task then would be to explain the
origin of the peculiar choice of SSB parameters at the GUT scale.

An attractive alternative has recently been suggested, and developed, in
a series of papers\cite{feng}. The idea is to start with multi-TeV
masses for {\it all} scalar fields at the GUT/Planck scale, and to
generate the IMH {\it radiatively} (RIMH). Neglecting sub-TeV
contributions to the renormalization group equations (RGE), and assuming
Yukawa couplings and also $A$ parameters remain unified at all mass scales, the
one loop RGEs for the third generation plus Higgs scalar SSB masses can
be solved explicitly. Remarkably, for simple forms of GUT scale boundary
conditions that are consistent with $SO(10)$ grand unification, Yukawa
couplings rapidly drive third generation SSB and Higgs boson masses
towards zero at the weak scale, while the first and second generation
scalars (which have negligible Yukawa couplings) remain in the multi-TeV
range. The contribution from the evolution of the gauge singlet right
handed neutrino (RHN) superfield $\hat{N}^c$ that occurs in $SO(10)$
unified theories plays a crucial role.  A crunch factor $S$ ---
essentially the average of first generation squared scalar masses
compared to third generation ones --- was defined.  Using GUT scale
Yukawa couplings of $f_{GUT}\sim 1-2$, and scale choices $M_N\sim
10^4-10^{15}$ GeV at which the RHN decouples from the MSSM, crunch
factors of $S\sim 50-800$ were claimed. As an example, a crunch factor
$S=400$ would imply that 10 TeV first and second generation scalars
could co-exist with 0.5 TeV third generation scalars at the weak scale.

A number of problems arise if one attempts a realistic implementation of
the RIMH model. The first, as noted also by the authors of
Ref. \cite{feng}, is that for models with large $\tan\beta$, which is
where Yukawa coupling unification occurs, it is difficult to generate a
radiative breakdown in electroweak symmetry (REWSB).  The second, as we
will see below, is that for model parameters which lead to large values
of $S$, the experimental values for third generation fermion masses are
not obtained.\footnote{Yukawa couplings, and hence fermion masses, for
the first two generations, and CKM matrix parameters, are neglected in Ref
\cite{feng}, in Ref. \cite{bmt}, and in this paper.}  A third problem
occurs in a full implementation of the renormalization group solution to
the RIMH model. The neglected sub-TeV mass contributions, two-loop RGE
contributions, and non-universal Yukawa couplings and $A$ parameters
below the GUT scale all serve to perturb the simple solutions found in
Ref. \cite{feng}, generally resulting in a considerable diminution of
the crunch factors that can be obtained.

It has been shown\cite{bdft} that the problem of implementing
REWSB in Yukawa unified $SO(10)$ models can be solved by including allowed
$D$-term contributions to scalar field masses: these occur when spontaneous
gauge symmetry breaking leads to a reduction in rank of the
gauge group\cite{km}.  The
form of the $SO(10)$ $D$-terms allows for a split in the GUT scale Higgs
masses so that $m_{H_u}^2<m_{H_d}^2$ which, in turn, facilitates
REWSB. The $D$-terms also affect
other scalar masses, resulting in a sparticle mass spectrum where the
bottom squark $\tb_1$ is frequently the lightest of all the matter
scalars, and left- sleptons are lighter than right- sleptons, as opposed
to the situation in the mSUGRA model, with GUT scale universality\cite{bdft}.

In a recent paper\cite{bmt}, it was shown that $SO(10)$ $D$-terms
could rescue REWSB into the RIMH framework as well,
so that sparticle masses could be explicitly calculated.
In Ref.\cite{bmt}, a bottom up approach was used to calculate
the values of GUT scale Yukawa couplings, and the
GUT scale SSB boundary conditions identified in Ref. \cite{feng} were adopted.
The sub-TeV contributions to RGEs were also included. The crunch factors
obtained, however, were in the range $S\sim 2-7$, much smaller than
the values found previously\cite{feng}.

In this paper, we perform a more detailed analysis of RIMH models. We
have two main goals. First, we seek a realistic implementation of RIMH
models, keeping track of sub-TeV and two-loop contributions to
RG evolution, requiring appropriate values of third generation fermion masses, and requiring a radiative breakdown in electroweak
gauge symmetry. We find it possible to construct realistic RIMH models,
but with crunch factors
$S \alt 10$. Thus, we conclude that realistic RIMH models give only a
partial solution to the SUSY flavor and CP problems. The RIMH model,
coupled with a limited degree of non-universality of soft breaking
scalar masses or alignment of fermion and sfermion matrices may be
enough to obtain a complete solution.
Once realistic SUSY particle mass spectra can be calculated within the
RIMH model context, our second goal is to examine various
phenomenological implications of these models, including
the neutralino relic density, $b\to s\gamma$ decay rates,
and expectations for collider searches.

Toward these ends, we upgrade the supersymmetric RG solution used in
ISAJET to include the complete set of two-loop RGEs for both the MSSM
and MSSM+RHN models. As first shown in Ref. \cite{am}, multi-TeV two
loop contributions can potentially be as large as sub-TeV leading order
contributions to the RGEs.  In Sec. II{\it a}, we first re-analyze RIMH
models using a bottom-up approach to the SUSY RG solution. The bottom-up
approach guarantees solutions with correct third generation fermion
masses. Allowing more flexibility in the value of $M_N$ allows somewhat
greater values (compared to Ref. \cite{bmt}) of $S\sim 9$ to be
obtained. These $S$ values are far below the values of $S$ claimed in
Ref. \cite{feng}. To understand this, in Sec. II{\it b}, we repeat much
of our analysis using a top-down RGE approach. In this case, large GUT
scale Yukawa couplings can be used. We find then that a significantly
larger IMH mass gap, with $S\sim 35$ can be obtained. However, in this
case the third generation fermion masses do not match their measured
values. In particular, the fixed point prediction of the value of $m_t$
is quite far from its experimental value.
The remainder of the paper focuses upon the phenomenological
implications of the model.  In Sec. III, we attempt to systematize the
parameter space explored in Sec. II, and show plots of sparticle masses
versus model parameters.  In Sec. IV, we present our
predictions for the cosmological neutralino relic density,
while those for the branching ratio of the $b\to s\gamma$ decay are
shown in Sec.~V.
Finally, in Sec. VI, we show the reach of the Fermilab Tevatron and CERN LHC
for RIMH models. In spite of the decoupling solution with multi-TeV
first and second generation scalars, a considerable reach is obtained
at the CERN LHC,
corresponding to values of $m_{\tg}\sim 1600-1700$ GeV,
assuming just 10 fb$^{-1}$
of integrated luminosity. We end with a summary of our results along
with some conclusions in Sec.~VII.
In an appendix, we list the two-loop RGEs for the
MSSM and MSSM+RHN models used in our analysis.

\section{RIMH model}

We begin by briefly reviewing the analysis of Bagger {\it et
al.}\cite{feng}.  Within the framework of the MSSM+RHN model (which is
assumed to be valid up to the GUT scale), the 1-loop RGEs for the third
generation and Higgs scalar masses and trilinear terms can be written as
\begin{equation}
\frac{d}{dt} {\bf m}^2 =-Y {\bf N} {\bf m}^2,
\label{eq:bfp1}
\end{equation}
where ${\bf m}^2=(m_{H_u}^2,
m_U^2,m_Q^2,m_D^2,m_{H_d}^2,m_L^2,m_E^2,m_N^2,A^2)^T$, $t$ is the
natural log of the energy scale and $Y={f^2\over{16\pi^2}}$.  In
Eq.~(\ref{eq:bfp1}), sub-TeV scale terms such as gaugino masses are
neglected.  Inspired by $SO(10)$ unification, $f$ here is the unified
Yukawa coupling, and $A$ is the unified trilinear SSB coupling and ${\bf
N}$ is a matrix of numerical coefficients obtained from the one-loop
RGEs. The qualitative behaviour of the weak scale SSB masses
can be obtained by neglecting the
splitting between the third generation Yukawa couplings and $A$
parameters that arises from renormalization group running.
The evolution of the SSB masses is then given by,
\begin{equation}
{\bf m}^2(t=t_f)=\sum c_i{\bf e}_i exp\left[-\lambda_i\int_0^{t_f} Y\ dt
\right] ,
\label{eq:bfp2}
\end{equation}
where $\lambda_i$ is the $i$th eigenvalue corresponding to
eigenvector ${\bf e}_i$ of the matrix ${\bf N}$, and $t_f$ is the
natural log of the weak scale.

The eigenvalues of the matrix ${\bf N}$ are non-negative, with the
largest eigenvalue $\lambda_1 =16$. Components of the initial conditions
along this direction are strongly suppressed by renormalization group
evolution to the weak scale: the corresponding
eigenvector suggests that the SSB masses should be related as
\begin{equation}
4m_Q^2=4m_U^2=4m_D^2=4m_L^2=4m_E^2=4m_N^2=2m_{H_u}^2=2m_{H_d}^2=A_0^2,
\label{eq:bfp3}
\end{equation}
at the GUT scale. This condition is consistent with minimal $SO(10)$
unification. Plugging the eigenvector ${\bf e}_1$ and eigenvalue
$\lambda_1$ into (\ref{eq:bfp2}), it is found that third generation SSB
masses are exponentially ``crunched'' to zero at the weak scale
$t_f$. First and second generation scalars have only tiny Yukawa
couplings, and hence, their masses are essentially  unchanged by the
evolution between the GUT and weak scales. This is the basic mechanism
for the radiative generation of the IMH.
The singlet neutrino superfield $N^c_3$,
which is automatically present in $SO(10)$ SUSY GUT models is critical
for this mechanism to work.

In our examination of the RIMH model, we adopt several assumptions motivated by
the work of Bagger {\it et al.}, Ref. \cite{feng}. First, we assume
the existence of an $SO(10)$ SUSY GUT model above the scale
$Q=M_{GUT}$, where $M_{GUT}$ is defined as the scale at which the
gauge couplings $g_1$ and $g_2$ unify. The $SO(10)$ gauge symmetry breaks to
$SU(3)_C\times SU(2)_L\times U(1)_Y$ at $M_{GUT}$, so that the
MSSM+RHN model is the effective theory immediately below the GUT scale.
The matter superfields of the model
include the usual superfields of the MSSM
plus a gauge singlet neutrino superfield:
\begin{eqnarray*}
\hat{Q}_i=\left( \begin{array}{c} \hat{u}_i \\ \hat{d}_i\end{array}\right) ,\
\hat{L}_i=\left( \begin{array}{c} \hat{\nu}_i \\ \hat{e}_i\end{array}\right) ,\
\hat{U}^c_i,\ \hat{D}^c_i,\ \hat{E}^c_i ,\ \hat{N}^c_i ,
\end{eqnarray*}
where $i=1,2,3$ corresponds to the various generations. In addition, the Higgs
multiplets are given by
\begin{eqnarray*}
\hat{H}_u({\bf 2})=\left( \begin{array}{c} \hat{h}_u^+ \\ \hat{h}_u^0
\end{array}\right) ,\
 {\rm and}\ \hat{H}_d({\bf \bar{2}})=
\left( \begin{array}{c} \hat{h}_d^- \\ \hat{h}_d^0
\end{array}\right) .
\end{eqnarray*}
The superpotential is given by
\begin{equation}
\hat{f}=\hat{f}_{MSSM}+f_{\nu}\epsilon_{ij}\hat{L}^i\hat{H}_u^j\hat{N}^c
+{1\over 2}M_N\hat{N}^c\hat{N}^c ,
\label{sprhn}
\end{equation}
where
\begin{eqnarray*}
 \hat{f}_{MSSM}=\mu\hat{H}_u^i\hat{H}_{di}+f_u\epsilon_{ij}\hat{Q}^i\hat{H}_u^j
\hat{U}^c+f_d\hat{Q}^i\hat{H}_{di}\hat{D}^c+f_e\hat{L}^i\hat{H}_{di}\hat{E}^c .
\end{eqnarray*}
$\epsilon_{ij}$ is the completely antisymmetric $SU(2)$ tensor
with $\epsilon_{12}=1$, and we neglect terms involving Yukawa couplings
of the first and second generation. Here, and in the following, we show
terms involving only third generation right handed neutrino and
sneutrino fields, and suppress the generation index on these.

The soft SUSY breaking terms can be written as
\begin{equation}
{\cal L} = {{\cal L}_{MSSM}}-m_{\tnu_R}^2 |\tnu_R |^2 +\left[
A_{\nu}f_{\nu}\epsilon_{ij}\tilde{L}^i H_u^j\tilde{\nu}_R^\dagger
+ \frac{1}{2} B_{\nu} M_N \tnu_R^2 + h.c. \right] ,
\label{LmssmN}
\end{equation}
where
\begin{eqnarray*}
{\cal L}_{MSSM}=-\sum_r m_r^2|\phi_r|^2&-&{1\over 2}\sum_\lambda
M_\lambda \bar{\lambda}_\alpha\lambda_\alpha +\left[ B\mu H_u^i H_{di} +
h.c \right ] \\ &+& \left[A_u
f_u\epsilon_{ij}\tQ^i H_u^j\tu_R^\dagger +A_d f_d\tQ^i H_{di}
\td_R^\dagger+A_e
f_e \tL^i\tH_{di}\te_R^\dagger + h.c. \right ] .
\end{eqnarray*}
The parameters $A_{\nu}$, $B_{\nu}$
and $m_{\tnu_R}$ are assumed to be comparable to the weak scale.
Below the scale $Q=M_N$, but above the weak scale, the RHN superfield is
integrated out, and
the MSSM is the effective theory.
In our renormalization group analysis, we use the full set of two loop
renormalization group equations for the MSSM and the MSSM+RHN models,
as given in the Appendix. These RGEs have been encoded into the
event generator ISAJET 7.51\cite{isajet}, which we use for this analysis.

\subsection{Bottom-up approach}

To generate supersymmetric spectra within the framework just described,
we adopt the bottom-up approach inherent in ISAJET. This ensures that
the matter fermion masses are as given by experimental
measurements. Beginning with weak scale central values of the three SM
gauge couplings and three third generation Yukawa couplings, we evolve
upwards in energy until $g_1=g_2$, where the GUT scale is identified. In
keeping with the spirit of $SO(10)$ unification, we then take the
various GUT scale SSB scalar masses to be
\begin{eqnarray*}
m_Q^2&=&m_E^2=m_U^2=m_{16}^2+M_D^2 \\
m_D^2&=&m_L^2=m_{16}^2-3M_D^2 \\
m_N^2&=&m_{16}^2+5M_D^2 \\
m_{H_{u,d}}^2&=&m_{10}^2\mp 2M_D^2 , \\
\end{eqnarray*}
where $m_{16}$ is a common mass of all scalars in the 16-dimensional
spinorial representation of $SO(10)$, and $m_{10}$ is the mass of the Higgs
scalars, which are assumed to reside in a single ten-dimensional
fundamental representation of $SO(10)$. The parameter $M_D$ reflects
the magnitude of the $D$-term contribution to scalar masses
induced by the breakdown of $SO(10)$; its magnitude is taken to
be a free parameter.
We also assume the gaugino masses and trilinear SSB
terms unify to $m_{1/2}$ and $A_0$ at $M_{GUT}$, respectively. We do not
impose Yukawa coupling unification at $M_{GUT}$, although as we shall
see, the largest IMHs are generated when Yukawa couplings most nearly
unify. Beginning at $M_{GUT}$, we evolve the SSB masses, gauge and
Yukawa couplings down in energy to a scale $Q=M_N$ using the full two
loop MSSM+RHN RGEs given in the Appendix of this paper. From $Q=M_N$ to
$M_{weak}$, we evolve according to two loop MSSM RGEs. At $M_{weak}$,
radiative electroweak symmetry breaking is imposed, using the
renormalization group improved one-loop scalar potential, and SUSY loop
corrections are included for the relation between matter fermion masses
and the corresponding Yukawa couplings\cite{pierce}.  An iterative
procedure is used in ISAJET to obtain the superparticle mass spectrum.

The model is completely specified by the parameter set,
\begin{eqnarray*}
m_{16},\ M_D^2,\ m_{1/2},\ M_N,\ \tan\beta ,\ sign(\mu ),
\end{eqnarray*}
where $m_{10}$ and $A_0$ are determined in terms of $m_{16}$ by the
boundary condition (\ref{eq:bfp3}).
We adopt pole masses $m_{\tau}=1.777$ GeV,
$m_b=4.9$ GeV and $m_t=175$ GeV.
We generate random samples of model parameters with the following
ranges:
\begin{eqnarray*}
500 &<&m_{16}<10000\ {\rm GeV},\\
0&<&M_D^2<m_{16}^2/3,\\
100&<&m_{1/2}<3000\ {\rm GeV},\\
10^{6} &<&M_N < 10^{16}\ {\rm GeV},\\
10&<&\tan\beta <55, \\
\mu &>&0\ {\rm or}\ \mu <0 .
\end{eqnarray*}

Our first results are shown in Fig. \ref{imh2_1}. Here, we show
the ``crunch'' factor $S$ is defined as,
\begin{eqnarray}
S   =
 {{3(m_{\tu_L}^2+m_{\td_L}^2+m_{\tu_R}^2+m_{\td_R}^2)+m_{\te_L}^2+m_{\te_R}^2+
m_{\tnu_e}^2}
 \over
{3(m_{\tst_1}^2+m_{\tb_1}^2+m_{\tst_2}^2+m_{\tb_2}^2)+m_{\ttau_1}^2+
m_{\ttau_2}^2+ m_{\tnu_{\tau}}^2}},
\label{crunch}
\end{eqnarray}
for solutions to the superparticle mass spectrum consistent with
REWSB. We take $\mu <0$ and $A_0 <0$ in this figure.  Our definition of
$S$ differs slightly from the one in Ref.\cite{feng} since we are able
to use mass eigenvalues in the calculation.  In frame {\it a}), we show
$S$ versus $m_{16}$.  We first see that the largest values of $S$
obtained are close to $\sim 7$. In dedicated searches, values of $S$ as
high as 9 have been generated. The highest $S$ values occur for
$m_{16}\sim 2000$ GeV. For higher values of $m_{16}$, two loop RGE can
become important, thus perturbing even more the idealized solution of
Ref. \cite{feng}, and possibly leading to tachyonic mass
parameters\cite{am,gsimh}.  In frame {\it b}), we show $S$ versus
$\tan\beta$. Here, it is easy to see that the greatest IMH develops for
very large values of $\tan\beta \sim 50$ where Yukawa unification can
occur. Frame {\it c}) shows $S$ versus a ratio $R$ indicative of the
degree of Yukawa coupling unification.  We define
$R_{tb}=max(f_t/f_b,f_b/f_t)$, with similar definitions for
$R_{t\tau}$ and $R_{\tau b}$, and where $f_t$, $f_b$ and $f_\tau$ are the
third generation Yukawa couplings evaluated at $Q=M_{GUT}$.  Then, $R$
is the maximum of $R_{\tau b}$, $R_{tb}$ and $R_{t\tau}$.  The largest
$S$ values occur at values of $R\sim 1$, showing that the largest
hierarchy is obtained for models with the best Yukawa coupling
unification.  Finally, in frame {\it d}), we plot model solutions versus
the superpotential neutrino mass $M_N$. In this case, there is a mild
correlation to obtain higher $S$ values at small to intermediate values
of $M_N$, although a significant IMH can be generated at any of the
$M_N$ values shown.

In Fig. \ref{imh2_2}, we plot in {\it a}) $S$ versus the ratio
$m_{1/2}/m_{16}$. In this case, we see $S$ reaching a maximum for
$m_{1/2}/m_{16}\sim 0.2$. If $m_{1/2}$ is comparable to $m_{16}$,
then the effect of gaugino masses will be important, and will destroy
the radiative IMH solution. If $m_{1/2}$ is too low, then would-be
large $S$ models are excluded because we require that the lightest
SUSY particle (LSP) should be a neutralino: in these scenarios, for a
fixed value of $m_{16}$, third generation sfermion masses decrease
faster with $m_{1/2}$ than does $m_{\tz_1}$.
In frame {\it b}), we show $S$ versus the ratio
$M_D/m_{16}$. The distribution is bounded from above by $1/\sqrt{3}$;
higher values imply tachyonic sbottom masses at the GUT scale.
We see that $S$ reaches a maximum at $M_D/m_{16}\sim 0.2$.
Models with a high value of $\tan\beta$ required by Yukawa
unification need a non-zero value of $M_D$ to achieve a
consistent REWSB mechanism; models with $M_D\sim 0$ are allowed in our
plot, but will typically have smaller $\tan\beta$ values, and hence
a smaller degree of Yukawa unification.

In Figs. \ref{imh2_3} and \ref{imh2_4}, we show similar plots, but for
$\mu >0$. Many of the results are qualitatively the same
as in Figs. \ref{imh2_1} and \ref{imh2_2}.
An exception occurs
in the $S$ versus $R$ plot, where we see that values of $R\simeq  1$
are not allowed. This is a reflection of the well-known lack of Yukawa coupling
unification for $\mu >0$ in a bottom-up approach.
We have also made similar plots for $A_0>0$. In general, these models lead
to smaller values of $S$, and so are less promising than the $A_0<0$
models. We do not show these for reasons of brevity.

Our conclusion is that the maximum crunch factors obtainable in
realistic RIMH models are typically $S\sim 7-9$. While these crunch
factors are considerably smaller than values found in Ref. \cite{feng},
they can still lead to models with multi-TeV first and second generation
scalar masses, which is sufficient to solve several, but not all,
of the problems associated with FCNCs and CP violating processes.

In Table \ref{tab:1}, we show two sample RIMH model solutions, taking
$m_{16}=2500$ GeV, $m_{1/2}=650$ GeV, $\tan\beta =50$, and $M_D=500$
GeV. The first solution, for $\mu <0$, has Yukawa coupling unification
to $\pm 1.5$\%, with GUT scale Yukawa couplings at $f\simeq 0.54$. The
$\tb_1$ squark is the lightest of the third generation scalars, with
$m_{\tb_1}=446.3$ GeV, while first and second generation scalars have
masses around $2400-2800$ GeV.  Note the relatively heavy light Higgs
scalar $m_h=129.1$ GeV, which may make detection at Fermilab Tevatron
experiments difficult. The tau sleptons have mass just beyond 1 TeV,
while the lighter top squark has $m_{\tst_1}=858.8$ GeV.

The second case study has all the same parameters,
but the sign of $\mu$ is flipped: $\mu >0$.
In this case, SUSY loop corrections to Yukawa couplings
make Yukawa unification difficult,
with $f_t(M_{GUT})=.495$, while $f_b(M_{GUT})=.305$.
The bottom
squarks are all beyond 1 TeV, but $m_{\tst_1}=742.0$ GeV, and the
light tau slepton is the lightest third generation scalar with
$m_{\ttau_1}=661.2$ GeV.
For both cases, first and second generation scalars are heavy enough
to suppress many flavor changing and $CP$-violating effects, but not all
of them. Such a model would need to be coupled to some limited universality
to solve, for instance, the SUSY flavor and CP problems in the kaon
sector.

\subsection{Top-down approach}

In the bottom-up approach of the previous subsection, starting with
central values for the weak scale Yukawa couplings, we only were able to
obtain GUT scale Yukawa couplings of order $f\sim 0.5$.
These values of GUT scale Yukawa couplings are significantly lower than
the choices used in Bagger {\it et al.}, where $f(M_{GUT})\sim 1.0-2.4$.
Extrapolating their results to lower values of GUT scale Yukawa couplings
indicates that $S \alt 25-30$, only about a factor 3 larger than what we
find. We attribute this difference to the additional effects such as
inclusion of sub-TeV gaugino mass terms and two loop effects in the RGEs,
as well as deviation from the ideal boundary condition due to the
$D$-term contributions to scalar masses, all of which would result in a
smaller crunch factor.

To examine why we were confined to small values of the unified Yukawa
coupling, we modified our computer code to adopt a top-down RGE solution
in ISAJET. In this case, we could use arbitrary values of a unified GUT
scale Yukawa coupling as inputs, and compute the masses of third
generation fermion (as well as their superpartners).  Our results for
$\mu <0$ are exhibited in Fig. \ref{tdmum}.  In frame {\it a}), we show
the pole mass $m_t$ versus the value of $f(M_{GUT})$.  The well-known
Yukawa coupling fixed point behavior\cite{fixpt} is evident in this
plot, where a large range of values of $f(M_{GUT})= 1-2$ leads to
$m_t\simeq 200-210$ GeV.  We see, however, that values of
$f(M_{GUT})\simeq 0.5$ are necessary in order to obtain $m_t=175$ GeV,
and further, that for $f(M_{GUT}) > 1$, $m_t \geq 190$~GeV, which is
conclusively excluded by experimental measurements.\footnote{The reader
might worry that we have used 1-loop relations for the connection
between pole and running top masses, or alternatively, that these might
be scheme dependent. We have checked that in the $\overline{MS}$ scheme,
the difference between the 1- and 3-loop\cite{topMS} running top masses
is $< 2.2$~GeV, and further, that the difference between the running
masses in the $\overline{MS}$ and $\overline{DR}$ schemes\cite{topDR} is
$< 2$~GeV. As stated previously, we have included 1-loop SUSY radiative
corrections\cite{pierce} to the relation between the Yukawa couplings
and the corresponding fermion masses, and found these to be significant.}

In frame {\it b}), we show the crunch factor $S$ as a function of
$f(M_{GUT})$. Taking $f(M_{GUT})\simeq 0.5$, which leads to the measured
central value of the top quark mass, yields $S\alt 5$. Taking much
larger values of $f(M_{GUT})$ can lead to a much greater IMH, with
$S\sim 25-35$ being possible for $f(M_{GUT}) \simeq 1.2$, to be compared
with $S \alt 100$ in Ref. \cite{feng}. However, such models will in
general lead to a value of the top quark mass significantly above its
measured value. 
For yet larger values of $f(M_{GUT})$, electroweak symmetry is not
correctly broken, and these large $S$ solutions are no longer obtained:
for solutions with $\tan\beta$ near the maximum of its allowed range, we
get $\mu^2 < 0$ when $f(M_{GUT})$ is increased; for other cases, we find
that third generation mass parameters become tachyonic.\footnote{The
precise value of $f(M_{GUT})$ for which this occurs is somewhat sensitive
to the procedure that we adopt for numerically integrating the RGEs. We
take all sfermion masses to be $m_{16}$ during the first iteration of the
RGEs. Especially in the case of the IMH model, this grossly overestimates
the scale $Q$ at which the Higgs potential parameters are frozen in
ISAJET. In some cases though this results in tachyonic third generation
masses at this stage, and the parameter point is rejected.  While this is
indeed an artifact of our calculational procedure, we do expect tachyonic
third genration mass parameters when the Yukawa coupling becomes
sufficiently large. Thus, while the precise value of $f(M_{GUT})$ beyond
which there are no large $S$ solutions may be somewhat different from that
shown in the figure, we believe that the figure is qualitatively correct.
Since the value of $f(M_{GUT})$ at which the large $S$ solutions really
cuts off is only a matter of academic interest, and has nothing to do with
the real phenomenology of the model, we have not pursued this any
further.}
Frame {\it c}) shows $S$ versus the ratio $M_D/m_{16}$
in the top-down approach. The largest $S$ values are obtained for
smaller values of the $SO(10)$ $D$-term. Larger $D$-terms not only
perturb the boundary conditions of Ref. \cite{feng}, but can lead to
tachyonic bottom squark masses. Finally, in frame {\it d}), we show $S$
versus the ratio $m_{1/2}/m_{16}$. In this case, there is a
tight correlation, with the largest $S$ values being obtained for small
values of $m_{1/2}$. Larger $m_{1/2}$ values again will increasingly
perturb the simple evolution described in Ref. \cite{feng}, which was
obtained assuming that gaugino
mass could be neglected.
We have also analyzed the corresponding situation for $\mu > 0$. The
results, which are shown in Fig.~\ref{tdmup}, frames {\it
a})--{\it d}) are very similar to those in Fig.~\ref{tdmum} just discussed.

We illustrate the spectrum of a model with a large unified GUT scale
Yukawa coupling which yields $S = 33.8$ in Table \ref{tab:2}.  Here, we
take $f(M_{GUT})=1.28$, with other parameters as shown. In this case,
first generation masses of $\sim 12.5$ TeV are generated, while third
generation scalar masses are typically $1-3$ TeV. The spectrum is meant
for illustrative purposes only: not only is the top quark mass for this
case $m_t=204.8$ GeV, but third generation scalars are too
heavy to claim that the weak scale is really stabilized.

\section{Sparticle masses in the RIMH model}

In this section, we attempt to systematize the
parameter space and map out sparticle masses in preparation for
a phenomenological analysis of the RIMH model.
>From Figs. \ref{imh2_1} and \ref{imh2_3}, we see that the largest
crunch factors are obtained for large $\tan\beta$ values of $\sim 50$,
where Yukawa coupling unification occurs with the best precision.
In this region of parameter space, $D$-term contributions
to scalar masses are needed to gain consistency with REWSB, and in fact
we see from Figs. \ref{imh2_2} and \ref{imh2_4} that values of
$M_D\sim 0.2 m_{16}$ lead to the largest crunch factors.

Motivated by these considerations, we show in Fig. \ref{imh2_5} the
allowed regions of the $m_{16}\ vs.\ m_{1/2}$ parameter plane, taking
$\tan\beta =50$, $M_D=0.2 m_{16}$, $M_N=1\times 10^7$ GeV and $\mu
<0$.\footnote{Our choice of $M_N$ means that the simple see-saw
mechanism (with a generation-independent RHN mass) will not give
neutrino masses in agreement with the Super Kamiokande data. There must
be some additional physics in the neutrino sector which, we presume, will
not significantly affect sparticle phenomenology.} Throughout the rest
of the paper we take $A_0 < 0$, since this generally allows a bigger
IMH. For the lattice of points in this plane we have used ISAJET to
obtain the sparticle masses, subject of course to several theoretical
constraints.  Points with triangles are excluded because the LSP is not
a neutralino, points with open circles are excluded because they do not
lead to REWSB ($\mu^2 < 0$), while those denoted by dots lead to
tachyonic scalar mass parameters.
We show contours of
regions where $S>3$ and $S>5$, indicating that $S$ increases as one
proceeds towards the lower-right region of the plot.  Values of $S$ as
high as 8 can be obtained adjacent to the lower right excluded region.
Contours of $m_{\tg}=2$ TeV and $m_{\tu_L}=3$ TeV are also shown, with
particle labels on the contour side with lower mass values. We also show
contours of $m_{\tst_1}$ and $m_{\tb_1}$ of 1 TeV, and $\mu =-1$
TeV. The regions below the $\tst_1$ and $\tb_1$ contours are most
natural, while the region above the $\tu_L$ contour best suppresses FCNC
and CP violating loop effects. Finally, a contour shows the region where
$R<1.05$, corresponding to Yukawa coupling unification within $\pm 2.5$\%.

To obtain an idea of how some of these quantities change with
$M_N$, we plot {\it a}) $S$,
{\it b}) $R$, {\it c}) $m_{\tb_1}$ and {\it d}) $m_{\tst_1}$
versus $m_{16}$ for $m_{1/2}=0.25 m_{16}$ in Fig. \ref{imh2_6} for
several values of $M_N$ indicated on the figure. The curves cut off when
they run into the excluded regions discussed in Fig.~\ref{imh2_5}.
>From {\it a}), we see that,
for a given value of $m_{16}$, the largest $S$ values are obtained for
the smallest values of $M_N$. From {\it b}), the smallest $M_N$
values also lead to models with the best Yukawa coupling unification.
In {\it c}), it is seen that low values of $m_{16}$ and $M_N$ can
lead to very small values of $m_{\tb_1}$, some of which might
even be accessible to direct $\tb_1\bar{\tb}_1$ search experiments at
the Fermilab Tevatron\cite{sbottom}. In {\it d}), the lightest
top squark masses are generated for small $m_{16}$ but larger values
of $M_N$, so that some parts of model parameter space may be
accessible to direct $\tst_1\bar{\tst}_1$ search experiments at
Fermilab Run 2\cite{stop}.

In Fig. \ref{imh2_mass1}, we show various sparticle and Higgs boson
masses and the $\mu$ parameter versus $m_{16}$ for the same parameters
as in Fig. \ref{imh2_6}, but with $M_N=1\times 10^7$ GeV.\footnote{ The
calculation of $\mu$ near the edge of the excluded region (where $|\mu
|\to 0$) is numerically delicate, so we have increased the number of
Runge-Kutta steps to 1000 and convergence limits to 0.002, beyond their
default ISAJET values, for these plots.}  In frame {\it a}), a
significant mass gap is seen between first and third
generation scalars. Although the value of $S$ decreases as $m_{16}$
increases, the mass gap between the two generations is increasing. For
the first generation sleptons, we see that $m_{\te_L}< m_{\te_R}$, as
opposed to the situation in mSUGRA models with universality of scalar
masses. This can be a result of both $D$-term contributions to scalar
masses, and the effect of two-loop contributions to the
RGEs\cite{gsimh}. In frame {\it b}), we see that while $m_{\tg}$ and
$m_A$ are increasing with $m_{16}$, the value of $|\mu |$ is decreasing,
and in fact where $\mu\to 0$ forms the upper limit of the $m_{16}$
parameter space. This impacts upon the nature of the two lighter
neutralinos and the light chargino, which are gaugino-like for small
$m_{16}$, but become increasingly higgsino-like for larger values of
$m_{16}$, and has implications for the cascade decay patterns\cite{bbkt}
of gluinos, which should be accessible\cite{lhc} at the LHC.

In Fig. \ref{imh2_7}, we show again the $m_{16}\ vs.\ m_{1/2}$ plane,
as in Fig. \ref{imh2_5}, but for $\mu >0$. In this case, much more of the
parameter plane is excluded. The measure of Yukawa unification $R$
varies from $1.6-1.9$ across the allowed region, and a slightly reduced
value of crunch factor $S$ is obtained compared to the negative $\mu$
case.
In Fig. \ref{imh2_8}, we show the corresponding values of
$S$, $R$, $m_{\tb_1}$ and $m_{\tst_1}$ versus $m_{16}$, with the same
parameters as in Fig. \ref{imh2_6}, except $m_{1/2}=0.22 m_{16}$, and
$\mu >0$. In this case, the largest $S$ values are obtained for the
largest values of $M_N$, which is also where better Yukawa unification is
achieved. The $\tb_1$ mass here is always rather high, and not accessible
to Tevatron searches. However, the $\tst_1$ is the lightest of the scalars,
and may be accessible to Tevatron searches at Run 2 if $m_{16}$ is small
enough. Its mass is almost independent of the value of $M_N$.
Finally, in Fig. \ref{imh2_mass2}, we show again various sparticle masses
for the same parameters as in Fig. \ref{imh2_8}, but with
$m_{1/2}=0.22m_{16}$ GeV. Once again, we see that the light chargino and
the two lighter neutralinos go from being gaugino-like to higgsino-like
as $m_{16}$ increases.

Before moving on to the phenomenology, we remark that several authors
have suggested that since $\mu$ enters the tree level expression that
determines $M_Z$ via the REWSB constraint, $\mu \alt 300$~GeV might
provide a more valid determination of naturalness than the third
generation scalar masses. Adopting this criterion, we would then
conclude that in the region of the parameter plane in Fig.~\ref{imh2_5}
or Fig.~\ref{imh2_7} close to the open circles (where $\mu^2 \rightarrow
0$) the theory would be technically natural. We note, however, that
$\mu$ generically tends to be large, and very sharply dives to zero over
a limited range of parameters. This means that a small change in, for
instance, $m_{16}$ could lead to quite a different value of $\mu$ in
this region. For this reason we do not adopt $\mu$ as a naturalness
measure in our analysis.

\section{Neutralino relic density}

The calculation of the cosmological relic density of neutralinos
$\Omega_{\tz_1}h^2$ provides an important constraint on supersymmetric models
with $R$-parity conservation.
$\Omega_{\tz_1}h^2 >0.02$ to account at least for galactic
rotation curves.
Analysis of the cosmic microwave background suggests a total energy
density of the universe $\Omega = 1.0 \pm 0.2$. With the Hubble
parameter estimated to be about 65 km/sec/Mpc, we have $\Omega h^2
\simeq 0.42$.
In a universe with 5\% each of baryonic matter and massive neutrinos, and a
cosmological constant with $\Omega_\Lambda h^2 \sim 0.25$, we would
expect $\Omega_{\tz_1} h^2 \sim 0.13$.  As an extreme, assuming no
contribution to dark matter from the cosmological constant, we would obtain
$\Omega_{\tz_1} h^2 \sim 0.38$.

Our estimate of the relic density of neutralinos in
the RIMH model follows the calculational procedure outlined
in Ref. \cite{bb_relic}. Briefly, we evaluate all tree level neutralino
annihilation diagrams exactly as helicity amplitudes. We then calculate
the neutralino annihilation cross section, and compute the thermally
averaged cross section times velocity using the fully relativistic
formulae of Gondolo and Gelmini\cite{gg}. Once the freeze-out
temperature is obtained via an iterative solution, we can
straightforwardly obtain the neutralino relic density
$\Omega_{\tz_1}h^2$. Our program takes special care to integrate
properly over any Breit-Wigner poles in $s$-channel annihilation
diagrams\cite{bb_relic}, an important feature to reliably
estimate the relic density at large values of the parameter
$\tan\beta$. We neglect co-annihilation effects, which can be important
when the $\tz_1$ is nearly mass degenerate with any of the other
sparticles\cite{coann}.

Our results for the neutralino relic density are shown in
Fig. \ref{imh2_rd}. We plot $\Omega_{\tz_1}h^2$ versus $m_{16}$
for both signs of $\mu$.
Here, we take
$\tan\beta =50$ and $M_D=0.2 m_{16}$. For $\mu <0$, we take
$m_{1/2}=0.25 m_{16}$, while for $\mu >0$, we take $m_{1/2}=0.22 m_{16}$,
{\it i.e.} we calculate along a diagonal strip in each of Figs.
\ref{imh2_5} and \ref{imh2_7}, in the region where $S$ is large.
For $\mu < 0$, values of $m_{16} \alt 2200$~GeV are in the excluded
region as can be seen from Fig.~\ref{imh2_5}.

Usually, in the mSUGRA model with large values for scalars masses at the
GUT scale, the relic density is too large as a result of scalar mass
suppression of the neutralino annihilation cross section. In fact we see
from Fig. \ref{imh2_rd} that significant ranges of $m_{16}$ can give
rise to either a too small or a too large neutralino relic density. In
the case that $\mu <0$, only a very narrow region with
$m_{16}\sim 4400-4600$
GeV gives $0.13 \alt \Omega_{\tz_1}h^2 \alt 0.38$, the favoured
range. In this region, $\tz_1$ is a mixture of higgsino and the bino, and
annihilates efficiently via $Z$ exchange, until for very large $m_{16}$
the LSP is dominantly a higgsino, and
the relic density becomes too small.\footnote{ Neutralino
co-annihilation will further diminish the relic density in this
region. This may slightly shift the allowed range of $m_{16}$ from that in
the figure.}  Conversely, for $m_{16}$ smaller than 4400~GeV, the LSP is
dominantly bino-like (see Fig.~\ref{imh2_mass1}{\it b}),
and the relic density rapidly increases because scalars are heavy and
the LSP couplings to gauge particles are suppressed. For yet smaller
values of $m_{16}$, the curve turns over and the relic density again
drops, mainly because of efficient annihilation into bottom pairs via
the exchange of a light $\tb_1$ (see Fig.~\ref{imh2_mass1}{\it a}) in
the $t$- and $u$-channels.
%
In the $\mu >0$ case, our relic density computation favours the region
around $m_{16}=3600$~GeV. For larger values of $m_{16}$, the relic
density again becomes too low for the same reason as for the $\mu < 0$
case. For smaller values of $m_{16}$, the relic density rapidly becomes
too large, but again turns over for $m_{16}< 3000$~GeV when the bottom
and top squarks are light enough to allow significant LSP annihilation to
$b\bar{b}$ and $t\bar{t}$ pairs, respectively. Indeed, the additional
turnover in the curve just below $m_{16}=2000$~GeV is exactly where
$\tz_1$ annihilation into top pairs becomes possible.
In summary, the relic
density constraints favour a relatively narrow range of $m_{16}$ where
the LSP is in the bino-higgsino transition region.  Furthermore,
significant portions of parameter space in the RIMH model, yield $\Omega
h^2 \ge 1$, and so lead to too young a universe, and are excluded.
However, viable parameter space regions can remain when certain squark
masses are sufficiently light, or where the $\tz_1$ is in the transition
region between being bino-like or higgsino-like.

We remark that in the parameter space region with a higgsino-like LSP,
it is important to note that rates for neutralino-nucleon scattering are
also enhanced\cite{bb_relic}, and this enhancement is compounded at
large $\tan\beta$. Thus, in the higgsino-like LSP region (where the
cosmogical dark matter must have a different origin), direct dark matter
search experiments can hope to find the first evidence of supersymmetry
by detection of neutralino-nucleon scattering.

\section{The decay $\lowercase{b} \to \lowercase{s} 
\gamma$ in the RIMH framework}

It is well known that the decay $b \to s\gamma$ provides a stringent
test of physics beyond the SM. The reason is that even within the SM
framework, this decay can only occur at the 1-loop level, via a loop
involving a top quark and a $W$ boson, with the photon attached to
either one, and so, the SM amplitude for this is often comparable in
magnitude to the corresponding amplitude from new physics.
The CLEO experiment\cite{cleo} still provides the best determination of
its branching fraction,
\begin{equation}
BR(b\to s\gamma )=(3.15\pm 0.35\pm 0.32\pm 0.26)\times 10^{-4}
\end{equation}
and restricts the branching ratio at 95\% CL to be
\begin{equation}
2\times 10^{-4}<BR(b\to s\gamma )<4.5\times 10^{-4} ,
\label{bsglim}
\end{equation}
while the precision,
\begin{equation}
BR(b \to s\gamma) = (3.34 \pm 0.5 \pm 0.36 \pm 0.27) \times 10^{-4} ,
\end{equation}
from the BELLE experiment\cite{belle} is not much less. These results
are in excellent agreement with the SM prediction, which at NLL accuracy
\cite{bsg_sm} is,
\begin{equation}
BR(b\to s\gamma )=(3.28\pm 0.33)\times 10^{-4} .
\end{equation}

The calculation of the width for $b\to s\gamma$ decay proceeds by
calculating the loop interaction for $b\to s\gamma$ within any particular model
framework, {\it e.g.} the MSSM,
at some high mass scale $Q\sim M_W$, and then matching
to an effective theory Hamiltonian given by
\begin{equation}
H_{eff}=-{4G_F\over \sqrt{2}} V_{tb}V^*_{ts}\sum_{i=1}^8 C_i(Q )O_i(Q ),
\label{eq1}
\end{equation}
where the $C_i(Q )$ are Wilson coefficients evaluated at scale $Q$, and
the $O_i$ are a complete set of operators relevant for the process $b
\to s\gamma$. We have included all order QCD corrections via
renormalization group resummation of the leading logs which arise
from the disparity of the new physics scale and the scale $m_b$ relevant
to the decay. Our procedure has
been described in detail elsewhere\cite{bsg}, and we will not repeat it here.

However, to enable the reader to get some feel for our results, we
mention that within the MSSM, the Wilson
coefficients $C_7$ and $C_8$ receive additional (1-loop) contributions
$tH^-$, $\tq_i \tw_j$, $\tq_i \tg$ and $\tq_i \tz_j$ loops, but the
first two, by far, dominate at least within the mSUGRA framework. We
should think of these coefficients as being renormalized at a scale
close to the scale of sparticle masses.
As mentioned above, the coefficients have then to be evolved down to $Q=m_b$.
The amplitude for the decay is proportional to $C_7(m_b)$ which, in
turn, depends on $C_i(M_W)$ ($i=2,7,8$).

We now turn to the results of our computation of $BR(b \to s\gamma)$
within the RIMH framework. We have computed this branching fraction for
the same two ``large $S$'' slices as in Fig.~\ref{imh2_rd}. Our result is
shown in Fig.~\ref{figbsg}. We see that for the slice with negative
$\mu$, the branching ratio that we obtain is considerably higher than
the 95\%CL upper limit (\ref{bsglim}), except for the largest values of
$m_{16}$ in this figure. If we recall that large $S$ models with negative
$\mu$ tend to give better Yukawa coupling unification, the reader will
immediately recognize that the conflict between the experimental value
and our prediction for the $b \to s\gamma$ decay has been noted before:
models with the sign of $\mu$ that allows Yukawa coupling unification
tend to give large values for $BR(b \to s\gamma)$ \cite{car,bdft}. We
note though that for $m_{16} \sim 4600$~GeV, the value favored by our
analysis of the relic density, the branching ratio is just about 25\%
larger than the CLEO upper limit. It is possible to imagine additional
effects ({\it e.g.} squark mixings or SUSY phases) that could serve to
reduce it to an acceptable level.

For the slice with $\mu>0$, we see that the RIMH model is in excellent
agreement with experiment for $m_{16}\sim 1800-2000$~GeV. Unfortunately,
this region of $m_{16}$ seems to be disallowed by our computation of the
relic density from neutralinos in Fig.~\ref{imh2_rd}. For smaller values
of $m_{16}$, the branching ratio rises rapidly to even beyond its
maximum for the negative $\mu$ case, while for larger values of $\mu$,
it decreases to well below the experimental lower bound from
(\ref{bsglim}). An especially intriguing feature is the sharp dip near
$m_{16}=3700$~GeV.

To gain some understanding of the behaviour seen in Fig.~\ref{figbsg},
we have examined the dominant contributions to the Wilson coefficients
$C_7$ and $C_8$ (as mentioned $C_2$ does not get any SUSY correction so
that $C_2(M_W)\simeq 1$) as a function of $m_{16}$ for the same parameter
space slices. We show our results for the negative and positive $\mu$
cases in Figs.~\ref{bsgn} and Figs.~\ref{bsgp}, respectively. In frame
{\it a}), we show our computation of $C_7$ while the corresponding
computation for $C_8$ is shown in frame {\it b}).  The dashed lines show
the dominant individual components, while the solid lines labelled
$C_7(M_W)$ and $C_8(M_W)$ show the sum of all individual
contributions. Finally, we have also shown the result after evolution of
these coefficients down to the weak scale. As mentioned, $C_7(m_b)$ is
the important quantity for the computation of the rate for the decay $b
\to s\gamma$, and it is for this computation that $C_{2,8}(Q)$ are
needed.

The following points are worth noting.
\begin{enumerate}
\item Contributions from gluino and neutralino loops are negligible and
are not shown.

\item The contribution from the $H^{\pm}$ loop is small, and only weakly
dependent on $m_{16}$, because $m_{H^{\pm}}$ (which is close to $m_A$)
is always large. The SM contribution is, of course, independent of
$m_{16}$. Essentially all the dependence on $m_{16}$ comes from the
$\tst_i\tw_j$ contributions, which tend to grow as $m_{16}$ gets close to
its lower end, primarily because the top squarks tend to be relatively
light, as can be seen from Fig.~\ref{imh2_mass1} and Fig.~\ref{imh2_mass2}.

\item For $\mu < 0$, the SUSY contributions interfere constructively
with the SM contribution. Since the latter, by itself, gives good
agreement with the experimental decay rate, it is not surprising that
the SUSY model prediction for the rate is too large for this sign of
$\mu$.

\item The SUSY amplitudes, which for large $\tan\beta$ are proportional
to $A_t \mu \tan\beta$, reverse their sign if $\mu > 0$, and interfere
destructively with the SM amplitude. Indeed it is this that makes it
possible for the RIMH model to be in agreement with the experimental
value for $BR(b \to s\gamma)$. For this to be possible, the RIMH model
amplitude (aside from the QCD evolution effect) has to have about the same
magnitude (but opposite sign) as the amplitude in the SM.

\item We see that the chargino contribution, and thus the total
amplitude, varies much more for the positive $\mu$ case than for the
case with negative $\mu$. Presumably, this is because top squarks can get much
lighter when $\mu > 0$ as can be seen from Fig.~\ref{imh2_mass2}. This
also explains why the range for the prediction in Fig.~\ref{figbsg} is
much wider for the positive $\mu$ case. The sharp dip in the prediction
of the branching ratio occurs when $C_7(m_b)$ goes through zero.

\item The evolution of $C_7$ between $Q=M_W$ and $Q=m_b$ is much more
sizeable for the $\mu > 0$ case than for $\mu < 0$. Indeed we see that
this effect reduces $\Gamma(b \to s \gamma)$, which should be
proportional to $|C_7|^2$, quite substantially when $\mu > 0$.
\end{enumerate}

\section{Reach of Fermilab Tevatron and CERN LHC for RIMH models}

Supersymmetric models with an IMH spectrum for the superparticles
present unique problems and opportunities for detection at collider
experiments. Since scalars of the first two generations have multi-TeV
masses, they effectively decouple, and will be produced (if at all) with
only tiny cross sections. Charginos, neutralinos and gluinos may also be
too heavy, so that search experiments may be forced to focus on direct
detection of third generation scalars. In addition, third generation
sleptons are generally difficult to detect at hadron colliders, so that
searches may have to focus on direct production of top and/or bottom
squarks.

If $\mu <0$, then from Fig. \ref{imh2_mass1} we see that the light bottom
squark $\tb_1$ can be the lightest of all matter scalars. It has been shown
that pairs of bottom squarks may be directly detectable
at the Fermilab Tevatron $p\bar{p}$ collider if $m_{\tb_1}\alt 240$ GeV,
provided $\tb_1\to b\tz_1$ is the dominant decay mode, $m_{\tz_1}$ is not
too heavy, and sufficient integrated luminosity is
obtained\cite{sbottom,run2}.
In Fig. \ref{imh2_5}, we have denoted with stars
in the lower left corner the points in
parameter space where direct detection of bottom squark pairs
should be possible, assuming 25 fb$^{-1}$ of integrated
luminosity at the Fermilab Tevatron. The region below the
associated dashed contour shows where $m_{\tb_1}< 240$ GeV.

If $\mu >0$, for the allowed region shown in Fig. \ref{imh2_7}, bottom
squarks are always too heavy to be detectable at the Tevatron. However,
for some parameter space points, $m_{\tst_1}<210$ GeV, the putative
reach of Tevatron upgrades\cite{stop}.  For these cases, however, the
associated value of $m_{\tz_1}$ from Fig.~\ref{imh2_mass2}{\it b} exceeds
140~GeV so that the top squark decay products are too soft to yield an
observable signal at the Fermilab Tevatron\cite{stop,run2}.

We have also examined in a limited context the observability of signals
in the
RIMH model at the CERN LHC $pp$ collider.
In Ref. \cite{lhc}, the reach of the CERN LHC for SUSY particles
within the mSUGRA model has been examined. In that model,
values of $m_{\tg}\sim 1500-2000$ GeV are generally accessible
assuming just 10 fb$^{-1}$ of integrated luminosity.
The strategy was to select events with
\begin{itemize}
\item jet multiplicity, $n_{\rm jet} \geq 2$ (with $E_{T,{\rm jet}} >
100$~GeV),

\item transverse sphericity $S_T > 0.2$,

\item $E_T(j_1), \  E_T(j_2) \ > \ E_T^c$ and $\eslt > E_T^c$,

\end{itemize}
where the parameter $E_T^c$ is adjusted to
optimize the signal.
We classify the events by the multiplicity of {\it isolated}
leptons, and in the case
of dilepton events, we also distinguish between the
opposite sign (OS) and same sign (SS)
sample as these could have substantially different origins. For the
leptons we require

\begin{itemize}

\item $p_T(\ell) > 20$~GeV ($\ell=e$ or $\mu$) and $M_T(\ell,\eslt) > 100$~GeV
for the $1\ell$ signal, and

\item $p_T(\ell_1,\ell_2) > 20$~GeV for $n=2,3,\ldots$ lepton signals. We
do not impose any $p_T(\ell)> E_T^c$ requirement on the leptons.

\end{itemize}

In Fig. \ref{yili1}, we show the maximum value of signal cross section
($\sigma_S$) over square root of background cross section
($\sqrt{\sigma_B}$) in fb$^{1/2}$, where the ratio is maximized over
five choices of $E_T^c=100$, 200, 300, 400 and 500~GeV. The numeral at
each plot point corresponds to the value of $E_T^c$ which gives the
maximum value, {\it e.g.} 1 corresponds to $E_T^c=100$ GeV, {\it etc.}.
The SM background has been taken from Ref. \cite{lhc}.  As before, the
negative (positive) $\mu$ plot is made for a slice out of
Fig.~\ref{imh2_5} (Fig.~\ref{imh2_7}), where $m_{1/2}=0.25 m_{16}$
($m_{1/2}=0.22 m_{16}$) along which $S$ is large. The other parameters
are shown on the figure.
This plot can be used to obtain the SUSY reach of the LHC in each of the
channels shown, for any integrated luminosity and any statistical
significance of the signal.  The dashed line labelled 10 fb$^{-1}$ shows
the $5\sigma$ level for that integrated luminosity. In Fig.~\ref{yili1},
we have in addition required that
$\sigma_S \geq 0.5$~$fb$ -- if this is not the case for
the $E_T^c$ value that maximizes the statistical
significance, we show a lower value of $E_T^c$ in this plot.
We have also checked that for the range of $m_{16}$ values for which the
signal in any channel satisfies our ``$5\sigma$'' and ``$5\ events$'' criteria,
$\sigma_S/\sigma_B$ is automatically larger than 0.2.
We see from frame {\it a})
that for the chosen slice of the $\mu < 0$ parameter plane, models
$m_{16}\alt 2400$~GeV should yield observable signals at the CERN LHC
for just 10 fb$^{-1}$ of integrated luminosity. This corresponds to a
value of $m_{\tg}\sim 1450$ GeV, with $m_{\tb_1}\sim 450$ GeV being the
smallest of the squark masses.  The best significance is achieved in the
$0\ell$ and $1\ell$ channels.  The corresponding results are shown in
frame {\it b}) for the slice with $\mu >0$. In this case, $m_{16}\sim
2600$ GeV can be probed for the same integrated luminosity,
corresponding to $m_{\tg}\sim 1400$ GeV, and the lighter top
squark with $m_{\tst_1}\sim 625$~GeV is the lightest squark.  We emphasize that
the cuts used above were designed to pick out signals from the cascade
decays of heavy squarks and gluinos in the mSUGRA framework, and
not for the IMH model with a relatively light bottom or top
squark. Nevertheless, we felt that it is instructive to show these
results because general purpose searches similar to these will almost
certainly be the first to be carried out when LHC data become available.

The reach can be somewhat extended if one concentrates on signals
associated with direct top or bottom squark production.  As an attempt
in this direction, we require events with exactly two tagged $b$-jets
(with $b$-tag efficiency of 50\%), together with the following cuts:
\begin{itemize}
\item $\eslt >100$ GeV,
\item $p_T(b-jet(1))>100$ GeV,
\item $p_T(b-jet(2))>50$ GeV,
\item $\eslt +\sum E_T(QCD-jets) >1500$ GeV,
\end{itemize}
where QCD jets must have $p_T> 25$ GeV.  The SM background is assumed to
come dominantly from $t\bar{t}$ production, for which we find
$\sigma_B(t\bar{t})= 1.26$ fb.  Our signal cross sections after these
cuts are plotted in Fig. \ref{yili2}, for the same parameter plane
slices as in Fig. \ref{yili1}.  In this case, for just 10 fb$^{-1}$ of
integrated luminosity, the reach in frame {\it a}) for $\mu <0$ extends
to $m_{16}\sim 2850$ GeV, and in {\it b}) for $\mu >0$ to $m_{16}\sim
2950$ GeV.  This corresponds to a reach in $m_{\tg}\sim 1700$ and 1600
GeV, respectively.  We note that the rather hard cuts that we have made
(to eliminate the top background) will likely cut out the direct $\tst_1$
or $\tb_1$ SUSY signal if
the sbottom or stop mass is very small. In this case, dedicated searches
for these may be possible. It should also be kept in mind that within the RIMH
framework, $t$ or $b$ squarks may also be within the reach of future linear
electron positron colliders.

\section{Summary and conclusions}

An inverted scalar mass hierarchy has been suggested\cite{imh} as a possible
way to ameliorate the SUSY flavor and CP problems without destabilizing
the weak scale. A particularly attractive approach pioneered in
Ref.\cite{feng} is to generate this hierarchy dynamically. These authors
identified a simple set of $SO(10)$ symmetric boundary conditions for
the renormalization group evolution of the soft SUSY breaking parameters
of an MSSM +RHN model, and showed that the large Yukawa couplings for
third generation sparticles drove their masses to sub-TeV values, leaving
first and second generation sparticle masses in the multi-TeV range.
Our main goals in this paper were to examine the extent of the hierarchy
that could be generated within realistic RIMH
supersymmetric models, and to explore the
phenomenological consequences of these scenarios.

In contrast to values of the crunch factor (\ref{crunch}) in the range
$S=50-800$ reported in Ref.\cite{feng}, we found that a much more
limited IMH could be generated, with $S \leq 9$.  We found several
reasons for the difference.  First, in our bottom-up approach, starting
from the observed values of matter fermion masses, to calculate third
generation Yukawa couplings, only solutions with $f(M_{GUT})\sim 0.5$
could be obtained, to be compared with $f(M_{GUT})\sim 1-2$ used in
Ref. \cite{feng}. Using a top-down approach, we checked that such large
values of the GUT scale Yukawa coupling lead to $m_t\sim 195-210$ GeV,
which is well beyond its experimental value measured at Fermilab. We
also verified that for $f(M_{GUT}) \sim 1.2$, values of $S$ up to 35
could be obtained but, of course, with unacceptable values of $m_t$.
The second difference came from the inclusion of numerous perturbations
to the simple analytical solution of Ref. \cite{feng} which reduce the
hierarchy that can be obtained.  These include {\it i.}) SSB terms of
order the weak scale in the RGE, {\it ii.})  splitting between Yukawa
couplings and also $A$-parameters below the $GUT$ scale, {\it iii.})
two-loop contributions to renormalization group running, and {\it iv.})
inclusion of $SO(10)$ $D$-term contributions to scalar masses: these
contributions, which are essential\cite{bmt} for obtaining solutions
with REWSB, perturb the simple boundary conditions suggested in
Ref. \cite{feng}.

While these considerations make it clear why we are unable to find model
solutions where third generation scalars are an order of magnitude
lighter than their counterparts of the first two generations, we have
found realistic models with REWSB that give first and second generation
scalar masses of several TeV, at the same time maintaining sub-TeV third
generation scalar masses.

We attempted a systematic exposition of parameter space leading to RIMH
models. The parameter space for which the largest hierarchy is obtained
is characterized by large
$\tan\beta\sim 50$, $A_0 <0$, $M_D\sim 0.2 m_{16}$, and for any fixed
value of $m_{1/2}$, as large as possible values of $m_{16}$ as allowed
by REWSB (see Figs. \ref{imh2_5} and \ref{imh2_7}).  The RIMH solution
can be obtained for either sign of $\mu$; solutions with negative $\mu$
tend to lead to a somewhat larger hierarchy.

To analyze the phenomenological implications of RIMH models, we have
selected two slices of the parameter space which yield large $S$, one
for each sign of $\mu$. The corresponding sparticle masses are shown as
a function of $m_{16}$ in Fig.~\ref{imh2_mass1} for $\mu < 0$, and in
Fig.~\ref{imh2_mass2} for $\mu > 0$. The model predictions for the
neutralino relic density are shown in Fig.~\ref{imh2_rd}. We found that
a significant fraction of the model parameter space is excluded because
it leads to too young a universe. This is because sparticles are heavy
so that the LSP annihilation cross sections are generally suppressed.
The relic density prediction falls close to or within the experimentally
favoured region for models characterized by either a very light bottom
squark ($\mu < 0$), or by a lightest neutralino in the transition region
between being bino-like and higgsino-like (both signs of $\mu$).

We have also evaluated the branching fraction for the decay $b \to
s\gamma$ as a function of $m_{16}$ for these same parameter space
slices. Our result is shown in Fig.~\ref{figbsg}. We see that models
with negative $\mu$ yield too large a rate for this decay, and are
essentially excluded by the CLEO experiment\cite{cleo} at the 95\%CL,
except for the largest values of $m_{16}$ in the figure. We note, however,
that the branching fraction is $\leq 25$\% larger than this upper
limit as long $m_{16}> 4600$~GeV. Since it is possible that other
effects (squark mixing, SUSY phases) may modify the calculation of this
branching ratio, it may be worthwhile to exercise some caution before
definitively ruling out these scenarios. Models with
positive $\mu$ can yield agreement with the experimental measurements
for $m_{16} \sim 1800-2000$~GeV. Unfortunately, this range is
conclusively excluded by the relic density constraints, assuming that
the LSP is absolutely stable (as it is within our framework). Combining
the constraints from the $b\to s\gamma$ decay with those from the relic
density, we see that values of $m_{16} \sim 4600-5000$~GeV and $\mu <
0$ might lead to the most viable of these large $S$ scenarios.

Finally, in the previous section, we have examined some consequences of RIMH
models for collider searches.  For the Fermilab Tevatron, only very
small regions of model parameter space could be accessed via direct
searches for bottom squark pair production. At the CERN LHC collider,
the limited IMH obtained means that a complete decoupling of heavier
states does not take place, in contrast to what happens\cite{gsimh} in GSIMH
models.  By searching for $\eslt +jets$ events with at least
two tagged $b$-jets, values of $m_{\tg}\sim 1600-1700$ GeV could be
probed with just 10 fb$^{-1}$ of data. There are, however, potentially
viable regions of parameter space (including the region favored by the
combined constraint from the neutralino relic density and radiative $b$
decay) where finding supersymmetric signals could pose a much bigger
challenge.

%
\acknowledgments
We thank J. Feng and N. Polonsky for discussions.
This research was supported in part by the U.~S. Department of Energy
under contract numbers DE-FG02-97ER41022, DE-FG03-94ER40833 and
DE-FG02-95ER40896 and in part by the University of Wisconsin
Research Committee with funds granted by the Wisconsin Alumni
Research Foundation.
%
\bigskip
\appendix
\centerline{\bf Appendix: Two loop RGEs for the MSSM plus RHN model}
\bigskip
In this appendix, we give the relevant two loop RGEs for the MSSM plus
right handed neutrino model adopted in this paper. We augment the
field content of the MSSM by adding a singlet neutrino superfield
$\hat{N}^c_i$ for each generation $i=1,2,3$. In keeping with the basic
structure of $SO(10)$ GUT models, we will assume that only the
third generation neutrino Yukawa coupling is significant, and will
neglect terms including neutrino Yukawa couplings for the first two
generations.
The form of the superpotential and SSB terms given in Eqs. \ref{sprhn}
and \ref{LmssmN} are then sufficient to determine the form of the
renormalization group equations.

The two loop RGEs for the MSSM+RHN model can be extracted from
Ref. \cite{mv}. For the gauge couplings, we find
\begin{equation}
\frac{d}{dt}g_a = \frac{g_a^3}{16\pi^2}B_a^{(1)}+
\frac{g_a^3}{(16\pi^2)^2}\left[ \sum_{b=1}^{3} B_{ab}^{(2)}g_b^2-
\sum_{x=t,b,\tau ,\nu} C_a^x f_x^2\right] ,
\end{equation}
where $B_a^{(1)}$ and $B_{ab}^{(2)}$ are given in Ref. \cite{mv},
and
\begin{equation}
C_a^{t,b,\tau ,\nu}=\left(\begin{array}{cccc} {26\over 5} & {14\over 5} &
{18\over 5} & {6\over 5} \\
6 & 6 & 2 & 2\\ 4 & 4 & 0 & 0\end{array}\right) .
\end{equation}

The expression for the gaugino masses can be written in terms of the
same matrices of coefficients:
\begin{equation}
\frac{d}{dt}M_a = \frac{2g_a^2}{16\pi^2}B_a^{(1)}M_a +
\frac{2g_a^2}{(16\pi^2)^2}\left[\sum_{b=1}^{3} B_{ab}^{(2)}g_b^2
(M_a+M_b)+\sum_{x=t,b,\tau ,\nu}C_a^x(f_x^2A_x-M_a f_x^2 )
\right] .
\end{equation}

For the Yukawa couplings, we find
\begin{equation}
\frac{d}{dt} f_{t,b,\tau ,\nu}=\frac{1}{16\pi^2}\beta_{f_{t,b,\tau ,\nu}}^{(1)}
+\frac{1}{(16\pi^2)^2}\beta_{f_{t,b,\tau ,\nu}}^{(2)},
\end{equation}
with
\begin{eqnarray}
\beta_t^{(1)}&=& f_t\left[ 6f_t^2+f_b^2+f_\nu^2-{16\over 3}g_3^2-3g_2^2-
{13\over 15}g_1^2\right] ,\\
\beta_b^{(1)}&=& f_b\left[ f_t^2+6f_b^2+f_\tau^2-{16\over 3}g_3^2-3g_2^2-
{7\over 15}g_1^2\right] ,\\
\beta_\tau^{(1)}&=& f_\tau\left[ 3f_b^2+4f_\tau^2+f_\nu^2
-3g_2^2- {9\over 5}g_1^2\right] ,\\
\beta_\nu^{(1)}&=& f_\nu\left[ 3f_t^2+f_\tau^2+4f_\nu^2
-3g_2^2-{3\over 5}g_1^2\right] ,
\end{eqnarray}
and the two loop contributions are given by
\begin{eqnarray}
 \beta_t^{(2)}&=& f_t\left[
-22f_t^4-5f_b^4-3f_\nu^4
-5f_t^2f_b^2-3f_t^2f_\nu^2-f_b^2f_\tau^2-f_\nu^2f_\tau^2
+\left( {6\over 5}g_1^2+6g_2^2+16g_3^2\right)f_t^2
\right. \nonumber \\ && \left.
+{2\over 5}f_b^2 g_1^2+
\left( g_2^2+{136\over 45}g_3^2\right)g_1^2+8g_2^2g_3^2
-{16\over 9}g_3^4+{15\over 2}g_2^4+{2743\over 450}g_1^4
\right],\\
 \beta_b^{(2)}&=& f_b\left[
-5f_t^4-22f_b^4-3f_\tau^4
-5f_t^2f_b^2-f_t^2f_\nu^2-3f_b^2f_\tau^2-f_\nu^2f_\tau^2
+\left( {2\over 5}g_1^2+6g_2^2+16g_3^2\right)f_b^2
\right.\nonumber\\& &\left.
+{4\over 5}f_t^2 g_1^2+{6\over 5}f_\tau^2 g_1^2
+\left( g_2^2+{8\over 9}g_3^2\right)g_1^2+8g_2^2g_3^2
-{16\over 9}g_3^4+{15\over 2}g_2^4+{287\over 90}g_1^4
\right],\\
 \beta_\tau^{(2)}&=& f_\tau\left[
-9f_b^4-10f_\tau^4-3f_\nu^4
-3f_t^2f_b^2-3f_t^2f_\nu^2-9f_b^2f_\tau^2-3f_\nu^2f_\tau^2
+\left( 16g_3^2-{2\over 5}g_1^2\right) f_b^2
\right.\nonumber\\& &\left.
+\left({6\over 5}g_1^2+6g_2^2\right) f_\tau^2
+{9\over 5}g_2^2 g_1^2
+{15\over 2} g_2^4+{27\over 2}g_1^4
\right], \\
 \beta_\nu^{(2)}&=& f_\nu\left[
-9f_t^4-3f_\tau^4-10f_\nu^4
-3f_t^2f_b^2-9f_t^2f_\nu^2-3f_b^2f_\tau^2-3f_\nu^2f_\tau^2
+\left(16g_3^2+{4\over 5}g_1^2\right)f_t^2
\right.\nonumber\\& &\left.
+\left({6\over 5}g_1^2+6g_2^2\right)f_\nu^2+{6\over 5}f_\tau^2 g_1^2
+{9\over 5}g_2^2g_1^2
+{15\over 2}g_2^4+{207\over 50}g_1^4
\right] .
\end{eqnarray}

For the $A$-parameters, we find
\begin{equation}
\frac{d}{dt} A_{t,b,\tau ,\nu}=\frac{1}{16\pi^2}\beta_{A_{t,b,\tau ,\nu}}^{(1)}
+\frac{1}{(16\pi^2)^2}\beta_{A_{t,b,\tau ,\nu}}^{(2)},
\end{equation}
with
\begin{eqnarray}
\beta_{A_t}^{(1)}&=& 2\left[\sum c_ig_i^2M_i+6f_t^2A_t+f_b^2A_b +
f_\nu^2A_\nu\right],\\
\beta_{A_b}^{(1)}&=& 2\left[\sum c_i'g_i^2M_i+6f_b^2A_b+f_t^2A_t +
f_\tau^2A_\tau\right],\\
\beta_{A_\tau}^{(1)}&=& 2\left[\sum c_i''g_i^2M_i+3f_b^2A_b+4f_\tau^2A_\tau +
f_\nu^2A_\nu\right],\\
\beta_{A_\nu}^{(1)}&=& 2\left[\sum c_i'''g_i^2M_i+3f_t^2A_t+4f_\nu^2A_\nu +
f_\tau^2A_\tau\right],
\end{eqnarray}
where the $c_i$, $c_i'$ and $c_i''$ are given in Ref. \cite{bbo}, and
$c_i'''=({3\over 5},3,0)$. Also,
\begin{eqnarray}
\beta_{A_t}^{(2)}&=& -{2\over 225}M_1\left(2743 g_1^2+225g_2^2+680g_3^2+
90f_b^2+270f_t^2\right) g_1^2-2A_\tau\left( f_b^2+f_\nu^2\right) f_\tau^2
\nonumber\\
& &-2g_2^2M_2\left(g_1^2+15g_2^2+8g_3^2+6f_t^2\right) -{16\over 45}g_3^2 M_3
\left(17g_1^2+45g_2^2-20g_3^2+90f_t^2\right)\nonumber\\
& &+{2\over 5} A_tf_t^2\left( 6g_1^2+30g_2^2+80g_3^2-25f_b^2-220f_t^2-
15f_\nu^2\right)
\nonumber\\
& &+{2\over 5}A_b f_b^2\left(2g_1^2-50f_b^2-25f_t^2-
5f_\tau^2\right) -2A_\nu f_\nu^2\left(3f_t^2+6f_\nu^2+f_\tau^2\right),\\
\beta_{A_b}^{(2)}&=& -{2\over 45}M_1\left(287 g_1^2+45g_2^2+40g_3^2+
18f_b^2+36f_t^2+54f_\tau^2\right) g_1^2\nonumber\\
& &-2g_2^2M_2\left(g_1^2+15g_2^2+8g_3^2+6f_b^2\right)-{16\over 9}g_3^2 M_3
\left( g_1^2+9g_2^2-4g_3^2+18f_b^2\right)\nonumber\\
& &+{2\over 5}A_tf_t^2\left(4g_1^2-25f_b^2-50f_t^2-5f_\nu^2\right)
+{2\over 5}A_\tau f_\tau^2\left( 6g_1^2-15f_b^2-5f_\nu^2-30f_\tau^2\right)
\nonumber\\
& &+{2\over 5}A_bf_b^2\left( 2g_1^2+30g_2^2+80g_3^2-220f_b^2-25f_t^2-
15f_\tau^2\right)-2A_\nu f_\nu^2\left(f_t^2+f_\tau^2\right) ,\\
\beta_{A_\tau}^{(2)}&=& -{2\over 5}M_1\left(135 g_1^2+9g_2^2
-2f_b^2+6f_\tau^2\right) g_1^2
-32g_3^2 M_3f_b^2-6A_tf_t^2\left(f_b^2+f_\nu^2\right)\nonumber\\
& &+{2\over 5}A_\tau f_\tau^2\left(6g_1^2+30g_2^2-45f_b^2-15f_\nu^2
-100f_\tau^2\right) -6A_\nu f_\nu^2\left(f_t^2+2f_\nu^2+f_\tau^2\right)
\nonumber\\
& &-{6\over 5}g_2^2M_2\left(3g_1^2+25g_2^2+10f_\tau^2\right)
-{2\over 5}A_b f_b^2\left( 2g_1^2-80g_3^2+90f_b^2+15f_t^2+45f_\tau^2\right),\\
\beta_{A_\nu}^{(2)}&=& -{2\over 25}M_1\left(207 g_1^2+45g_2^2
+20f_t^2+30f_\nu^2+30f_\tau^2\right) g_1^2
-32g_3^2 M_3f_t^2\nonumber\\
& &+{2\over 5}A_tf_t^2\left(4g_1^2+80g_3^2-15f_b^2-90f_t^2
-45f_\nu^2\right)-{6\over 5}g_2^2M_2\left(3g_1^2+25g_2^2+10f_\nu^2\right)
\nonumber\\
& &+{2\over 5}A_\nu f_\nu^2\left(6g_1^2+30g_2^2-45f_t^2-100f_\nu^2-
15f_\tau^2\right)+{6\over 5}A_\tau f_\tau^2\left(2g_1^2-5f_b^2-5f_\nu^2
-10f_\tau^2\right)\nonumber\\
& &-6A_bf_b^2\left(f_t^2+f_\tau^2\right) .
\end{eqnarray}

For the soft SUSY breaking scalar masses, we find
\begin{equation}
\frac{d}{dt}m^2=\frac{1}{16\pi^2}\beta_{m^2}^{(1)}+\frac{1}{(16\pi^2)^2}
\beta_{m^2}^{(2)},
\end{equation}
where for the Higgs masses, we have
\begin{eqnarray}
\beta_{m_{H_u}^2}^{(1)}&=&
6f_t^2X_t+2f_\nu^2X_\nu
-{6\over 5}g_1^2M_1^2-6g_2^2M_2^2
+{3\over 5}g_1^2{\cal S},\\
\beta_{m_{H_d}^2}^{(1)}&=&
6f_b^2X_b+2f_\tau^2X_\tau
-{6\over 5}g_1^2M_1^2-6g_2^2M_2^2
-{3\over 5}g_1^2{\cal S} ,
\end{eqnarray}
and
\begin{eqnarray}
\beta_{m_{H_u}^2}^{(2)}&=&
36{\cal A}_t + 12{\cal A}_\nu + 6{\cal B}_{bt} + 2{\cal B}_{\nu\tau}
+ \frac{8}{5}{\cal C}_{1t} + 32{\cal C}_{3t}
+{621\over 25} g_1^4M_1^2
+33g_2^4M_2^2
\nonumber\\ & &
+{18\over 5}{\cal M}_{12}
+{6\over 5}g_1^2{\cal S}'
+3g_2^2\sigma_2+{3\over 5}g_1^2\sigma_1 ,\\
\beta_{m_{H_d}^2}^{(2)}&=&
36{\cal A}_b + 12{\cal A}_\tau + 6{\cal B}_{bt} + 2{\cal B}_{\nu\tau}
- \frac{4}{5}{\cal C}_{1b} + 32{\cal C}_{3b} + \frac{12}{5}{\cal C}_{1\tau}
+{621\over 25} g_1^4M_1^2
\nonumber\\ & &
+33g_2^4M_2^2
+{18\over 5}{\cal M}_{12}
-{6\over 5}g_1^2{\cal S}'
+3g_2^2\sigma_2+{3\over 5}g_1^2\sigma_1 ,
\end{eqnarray}
where
\begin{eqnarray}
{\cal A}_i&=&-(A_i^2+X_i)f_i^4, ~~~~~~~~~~~~~~~~~~~ i=t,b,\tau,\nu \\
{\cal B}_{ij}&=&-(2A_iA_j+X_i+X_j)f_i^2f_j^2, ~~~~ i,j=t,b,\tau,\nu \\
{\cal C}_{ij}&=&g_i^2(2M_i(M_i-A_j)+X_j)f_j^2, ~~~ i=1,2,3, ~ j=t,b,\tau,\nu \\
{\cal M}_{ij}&=&g_i^2g_j^2\left( M_i^2+M_j^2+M_iM_j\right),~~~i=1,2,3 \\
X_t&=& m_Q^2+m_U^2+m_{H_u}^2+A_t^2 ,\\
X_b&=& m_Q^2+m_D^2+m_{H_d}^2+A_b^2 ,\\
X_\tau &=& m_L^2+m_E^2+m_{H_d}^2+A_\tau^2 ,\\
X_\nu &=& m_L^2+m_{\tnr}^2+m_{H_u}^2+A_\nu^2 ,\\
{\cal S} &=& m_{H_u}^2-m_{H_d}^2+Tr \left[ {\bf m}_Q^2-{\bf m}_L^2
-2{\bf m}_U^2+{\bf m}_D^2+{\bf m}_E^2\right],\\
{\cal S}'&=& f_\tau^2 (-2m_E^2+m_{H_d}^2+m_L^2)+
f_b^2 (-2m_D^2+3m_{H_d}^2-m_Q^2) +
f_t^2 (4m_U^2-3m_{H_u}^2-m_Q^2)\nonumber\\
& &+
f_\nu^2 (m_L^2-m_{H_u}^2)+{6\over 5}g_1^2Tr({\bf m}_E^2)
-{3\over 10}(g_1^2+5g_2^2)\left(m_{H_d}^2-m_{H_u}^2+Tr({\bf m}_L^2)\right)
\nonumber\\ & &
+{2\over 15}(g_1^2+20g_3^2)Tr({\bf m}_D^2)
-{16\over 15}(g_1^2+5g_3^2)Tr({\bf m}_U^2)
\nonumber\\ & &
+{1\over 30}(g_1^2+45g_2^2+80g_3^2)Tr({\bf m}_Q^2),\\
\sigma_1 &=& {1\over 5}g_1^2\left\{ 3(m_{H_u}^2+m_{H_d}^2)
+Tr\left[ {\bf m}_Q^2+3{\bf m}_L^2+8{\bf m}_U^2+2{\bf m}_D^2+6{\bf m}_E^2
\right]\right\} ,\\
\sigma_2 &=& g_2^2\left\{ m_{H_u}^2+m_{H_d}^2+Tr\left[ 3{\bf m}_Q^2+
{\bf m}_L^2\right]\right\} ,\ \ {\rm and}  \\
\sigma_3 &=& g_3^2 Tr\left[ 2{\bf m}_Q^2+{\bf m}_U^2+{\bf m}_D^2 \right] .
\end{eqnarray}
In the above equations the sums over color and flavor are included,
and $Tr$ simply means a sum over generations.
The term ${\cal S}=0$ if universality is imposed, but can be
significant if scalar masses are non-universal.

The third generation matter scalar $\beta$-functions are given by
\begin{eqnarray}
\beta_{m_{Q}^2}^{(1)}&=&
2f_t^2X_t+2f_b^2X_b
-{2\over 15}g_1^2M_1^2-6g_2^2M_2^2-{32\over 3}g_3^2M_3^2
+{1\over 5}g_1^2{\cal S},\\
\beta_{m_{U}^2}^{(1)}&=&
4f_t^2X_t
-{32\over 15}g_1^2M_1^2-{32\over 3}g_3^2M_3^2
-{4\over 5}g_1^2{\cal S},\\
\beta_{m_{D}^2}^{(1)}&=&
4f_b^2X_b
-{8\over 15}g_1^2M_1^2-{32\over 3}g_3^2M_3^2
+{2\over 5}g_1^2{\cal S},\\
\beta_{m_{L}^2}^{(1)}&=&
2f_\tau^2X_\tau +2f_\nu^2 X_\nu
-{6\over 5}g_1^2M_1^2-6g_2^2M_2^2
-{3\over 5}g_1^2{\cal S},\\
\beta_{m_{E}^2}^{(1)}&=&
4f_\tau^2 X_\tau
-{24\over 5}g_1^2M_1^2
+{6\over 5}g_1^2{\cal S},\\
\beta_{m_{\tnr}^2}^{(1)}&=&4f_\nu^2 X_\nu ,
\end{eqnarray}
and the two loop contributions are
\begin{eqnarray}
 \beta_{m_{Q}^2}^{(2)}&=&
20({\cal A}_b+{\cal A}_t) + 2({\cal B}_{b\tau}+{\cal B}_{t\nu})
+\frac{4}{5}({\cal C}_{1b}+2{\cal C}_{1t})
+{199\over 75}g_1^4M_1^2+33g_2^4M_2^2-{128\over 3}g_3^4M_3^2
\nonumber\\ & &
+{2\over 5}{\cal M}_{12}+32{\cal M}_{23}+{32\over 45}{\cal M}_{13}
+{2\over 5}g_1^2{\cal S}'
+{1\over 15}g_1^2\sigma_1+3g_2^2\sigma_2+{16\over 3}g_3^2\sigma_3,\\
 \beta_{m_{U}^2}^{(2)}&=&
32 {\cal A}_t + 4({\cal B}_{b,t}+{\cal B}_{t,\nu})
+\frac{4}{5}(-{\cal C}_{1,t}+15{\cal C}_{2,t})
+{3424\over 75}g_1^4M_1^2-{128\over 3}g_3^4M_3^2
\nonumber \\ & &
+{512\over 45}{\cal M}_{13}-{8\over 5}g_1^2{\cal S}'
+{16\over 3}g_3^2\sigma_3
+{16\over 15}g_1^2\sigma_1,\\
 \beta_{m_{D}^2}^{(2)}&=&
32{\cal A}_b + 4({\cal B}_{b,t}+{\cal B}_{b,\tau})
+\frac{4}{5}({\cal C}_{1,b}+15{\cal C}_{2,b})
+{808\over 75}g_1^4M_1^2-{128\over 3}g_3^4M_3^2
\nonumber \\ & &
+{128\over 45}{\cal M}_{13}
+{4\over 5}g_1^2{\cal S}'
+{16\over 3}g_3^2\sigma_3+{4\over 15}g_1^2\sigma_1,\\
 \beta_{m_{L}^2}^{(2)}&=&
12({\cal A}_\nu+{\cal A}_\tau) + 6({\cal B}_{b,\tau}+{\cal B}_{t,\nu})
+\frac{12}{5}{\cal C}_{1,\tau}
+{621\over 25} g_1^4M_1^2+33g_2^4M_2^2
\nonumber\\& &
+{18\over 5}{\cal M}_{12}
-{6\over 5}g_1^2{\cal S}'
+3g_2^2\sigma_2+{3\over 5}g_1^2\sigma_1 ,\\
 \beta_{m_{E}^2}^{(2)}&=&
16{\cal A}_\tau + 4(3{\cal B}_{b,\tau}+{\cal B}_{\tau,\nu})
+\frac{12}{5}(-{\cal C}_{1,\tau}+5{\cal C}_{2,\tau})
+{2808\over 25}g_1^4M_1^2
+{12\over 5}g_1^2{\cal S}'
+{12\over 5}g_1^2\sigma_1,\\
 \beta_{m_{\tnr}^2}^{(2)}&=&
16{\cal A}_\nu + 4(3{\cal B}_{t,\nu}+{\cal B}_{\nu,\tau})
+\frac{12}{5}({\cal C}_{1,\nu}+5{\cal C}_{2,\nu}) .
\end{eqnarray}

For completeness, we also list the relevant $\beta$-functions for
the superpotential $\mu$ parameter and $M_N$, and the soft breaking
terms $B$ and $B_\nu$:
\begin{eqnarray}
\beta_\mu^{(1)}&=& \mu\left(-{3\over 5}g_1^2-3g_2^2+3f_b^2+3f_t^2
+f_\tau^2+f_\nu^2\right) ,\\
\beta_B^{(1)}&=& {6\over 5} g_1^2M_1+6g_2^2M_2+6f_t^2A_t+6f_b^2A_b
+2f_\tau^2A_\tau +2f_\nu^2A_\nu ,\\
\beta_{M_N}^{(1)}&=& 4M_Nf_\nu^2 ,\\
\beta_{B_\nu}^{(1)}&=& 8A_\nu f_\nu^2 ,
\end{eqnarray}
and
\begin{eqnarray}
\beta_\mu^{(2)}&=&\mu\left( {207\over 50}g_1^4 +{9\over 5}g_1^2g_2^2+
{6\over 5}f_\tau^2g_1^2 +{15\over 2} g_2^4 -9f_b^4 -9f_t^4
-3f_\nu^4-3f_\tau^4\right. \nonumber\\
& &\left. +\left(16g_3^2 -{2\over 5}g_1^2\right) f_b^2
-6f_b^2f_t^2 +\left({4\over 5}g_1^2+16g_3^2\right)f_t^2
-2f_\nu^2f_\tau^2\right) ,\\
\beta_B^{(2)}&=&-36A_tf_t^4-36A_bf_b^4-12A_\tau f_\tau ^4-12A_\nu f_\nu^4-
12(A_t+A_b)f_t^2f_b^2-4(A_\tau+A_\nu)f_\tau^2f_\nu^2 \nonumber\\
& &+\left( \frac{8}{5}(A_t-M_1)g_1^2+32(A_t-M_3)g_3^2\right)f_t^2+
\left(-\frac{4}{5}(A_b-M_1)g_1^2+32(A_b-M_3)g_3^2\right)f_b^2 \nonumber\\
& &+\frac{12}{5}(A_\tau-M_1)g_1^2f_\tau^2-
30g_2^4M_2-\frac{18}{5}g_1^2g_2^2(M_1+M_2)-\frac{414}{25}g_1^4M_1 ,\\
\beta_{M_N}^{(2)}&=& {4\over 5}M_N f_\nu^2\left( 3g_1^2+15g_2^2
-15f_t^2-10f_\nu^2-5f_\tau^2\right),\\
\beta_{B_\nu}^{(2)}&=&8f_\nu^2\left(\frac{3}{5}(A_\nu-M_1)g_1^2+3(A_\nu-M_2)g_2^2
-3(A_t+A_\nu)f_t^2-A_\nu (4f_\nu^2+f_\tau^2) - A_\tau f_\tau^2 \right).
\end{eqnarray}
%
%

\newpage
%
%

\iftightenlines\else\newpage\fi
\iftightenlines\global\firstfigfalse\fi
\def\dofig#1#2{\epsfxsize=#1\centerline{\epsfbox{#2}}}

\begin{table}
\begin{center}
\caption{Weak scale sparticle masses and parameters (GeV) for two IMH model
case studies. The first has $\mu <0$, and the second has $\mu >0$.}
\bigskip
\begin{tabular}{lcc}
\hline
parameter & case 1 & case 2 \\
\hline
$m_{16}$ & 2500.0  & 2500.0 \\
$m_{10}$ & 3535.5 & 3535.5 \\
$M_D$    & 500.0 & 500.0 \\
$m_{1/2}$ & 650.0 & 650.0 \\
$M_N$ & $1.0\times 10^{7}$ & $1.0\times 10^{7}$ \\
$A_0$ & -5000.0 & -5000.0 \\
$\tan\beta$ & 50.0 & 50.0 \\
$m_{\tg}$ & 1547.4 & 1558.9 \\
$m_{\tu_L}$ & 2789.4 & 2784.2 \\
$m_{\td_R}$ & 2589.9 & 2585.5 \\
$m_{\tell_L}$ & 2367.7 & 2366.1 \\
$m_{\tell_R}$ & 2566.7 & 2567.1 \\
$m_{\tnu_{e}}$ & 2366.3 & 2364.8 \\
$m_{\tst_1}$& 858.8 & 742.0 \\
$m_{\tst_2}$& 1250.6 & 1500.4 \\
$m_{\tb_1}$ & 446.4 & 1422.0 \\
$m_{\tb_2}$ & 1126.0 & 1527.9 \\
$m_{\ttau_1}$ & 1089.0 & 661.2 \\
$m_{\ttau_2}$ & 1299.0 & 878.8 \\
$m_{\tnu_{\tau}}$ & 1088.5 & 862.4 \\
$m_{\tw_1}$ & 518.6 & 528.0 \\
$m_{\tz_2}$ & 518.5 & 527.8 \\
$m_{\tz_1}$ & 281.5 & 281.2 \\
$m_h$ & 129.1 & 129.1 \\
$m_A$ & 819.3 & 1839.9 \\
$m_{H^+}$ & 826.3 & 1844.2 \\
$\mu$ & -651.5 & 849.1 \\
$R$ & 1.03 & 1.74 \\
$f_t(M_{GUT})$ & 0.535 & 0.495 \\
$f_b(M_{GUT})$ & 0.544 & 0.305 \\
$f_\tau (M_{GUT})$ & 0.552 & 0.531 \\
$S$ & 7.0 & 4.6
\label{tab:1}
\end{tabular}
\end{center}
\end{table}
\begin{table}
\begin{center}
\caption{Weak scale sparticle masses and parameters (GeV) for an IMH model
using a top-down approach. In this case, the unified Yukawa coupling at the
GUT scale is $f(M_{GUT})=1.282$, and the top quark pole mass
is found to be $m_t=204.8$ GeV. We take $\mu <0$.}
\bigskip
\begin{tabular}{lc}
\hline
parameter & value \\
\hline
$m_{16}$ & 12766.3 \\
$m_{10}$ & 18054.3 \\
$M_D$    & 765.4 \\
$m_{1/2}$ & 732.3 \\
$M_N$ & $3.5\times 10^{8}$ \\
$A_0$ & -25532.7 \\
$\tan\beta$ & 52.7 \\
$m_{\tg}$ & 2052.3 \\
$m_{\tu_L}$ & 12684.7 \\
$m_{\td_R}$ & 12621.8 \\
$m_{\tell_L}$ & 12662.5 \\
$m_{\tell_R}$ & 12779.2 \\
$m_{\tnu_{e}}$ & 12662.3 \\
$m_{\tst_1}$& 1419.0 \\
$m_{\tst_2}$& 1955.5 \\
$m_{\tb_1}$ & 1041.3 \\
$m_{\tb_2}$ & 1931.7 \\
$m_{\ttau_1}$ & 3381.7 \\
$m_{\ttau_2}$ & 3725.7 \\
$m_{\tnu_{\tau}}$ & 3721.8 \\
$m_{\tw_1}$ & 735.7 \\
$m_{\tz_2}$ & 735.6 \\
$m_{\tz_1}$ & 376.4 \\
$m_h$ & 153.6 \\
$m_A$ & 779.0 \\
$m_{H^+}$ & 786.7 \\
$\mu$ & -2367.7 \\
$R$ & 1.0 \\
$S$ & 33.9
\label{tab:2}
\end{tabular}
\end{center}
\end{table}
\newpage
%

%
\begin{figure}
\dofig{5in}{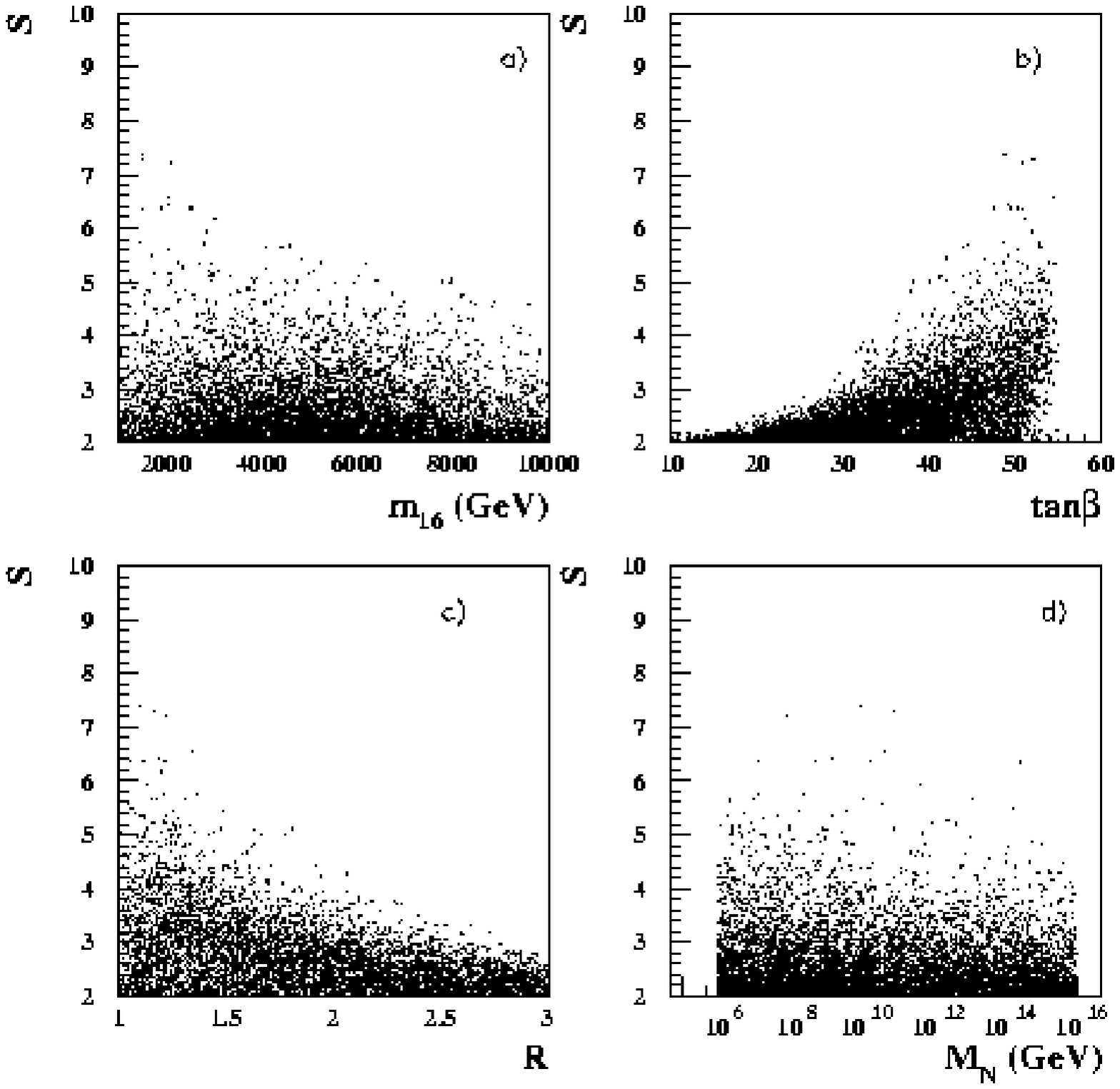}
\caption[]{
A plot of crunch factor $S$ versus {\it a}) $m_{16}$, {\it b}) $\tan\beta$,
{\it c}) $R$ and {\it d}) $M_N$, for all models with $A_0<0$ and $\mu <0$.}
\label{imh2_1}
\end{figure}
\begin{figure}
\dofig{5in}{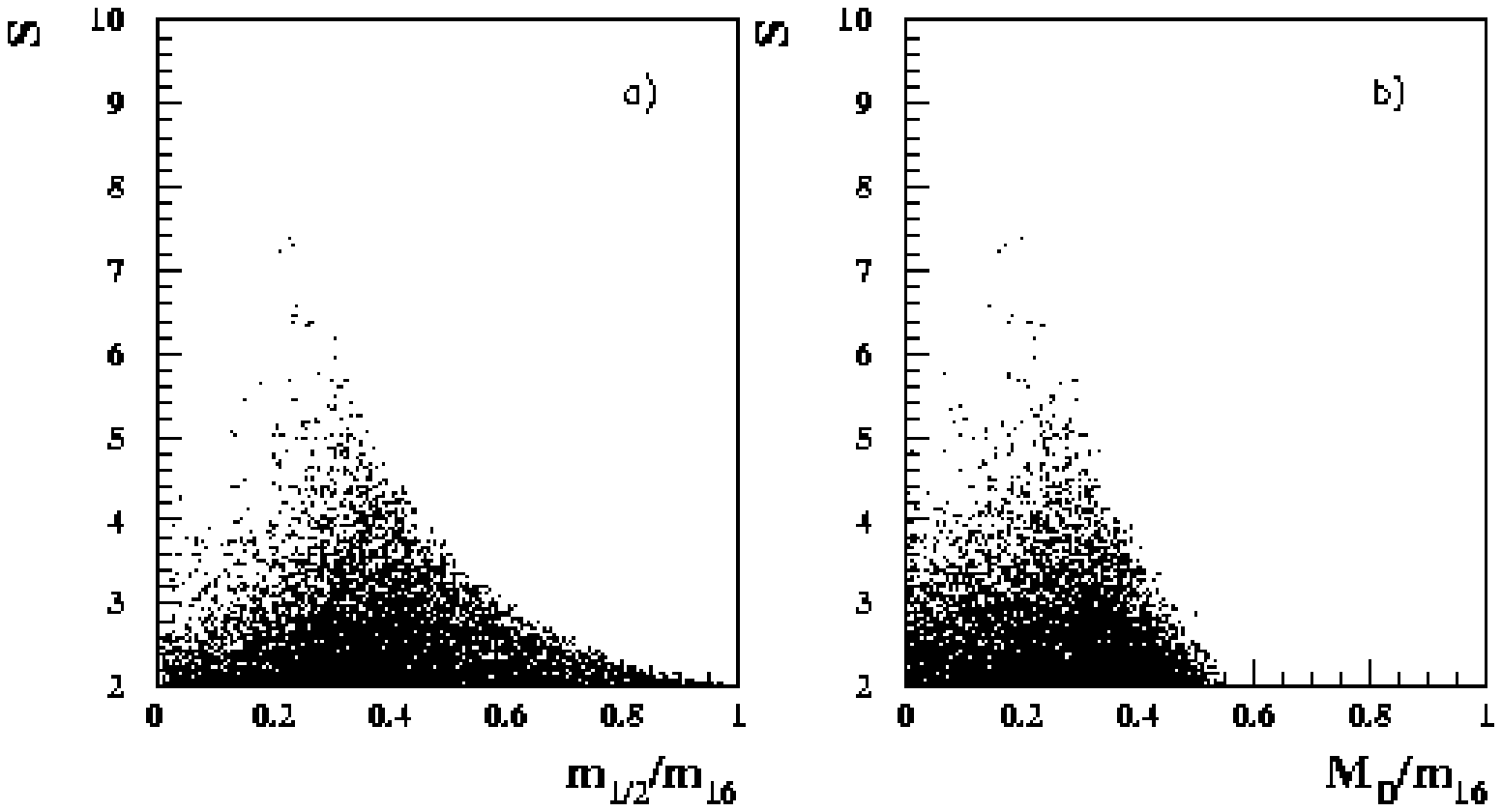}
\caption[]{
A plot of crunch factor $S$ versus {\it a}) $m_{1/2}/m_{16}$ and
{\it b}) $M_D/m_{16}$, for all models with $A_0<0$ and $\mu <0$.}
\label{imh2_2}
\end{figure}
\begin{figure}
\dofig{5in}{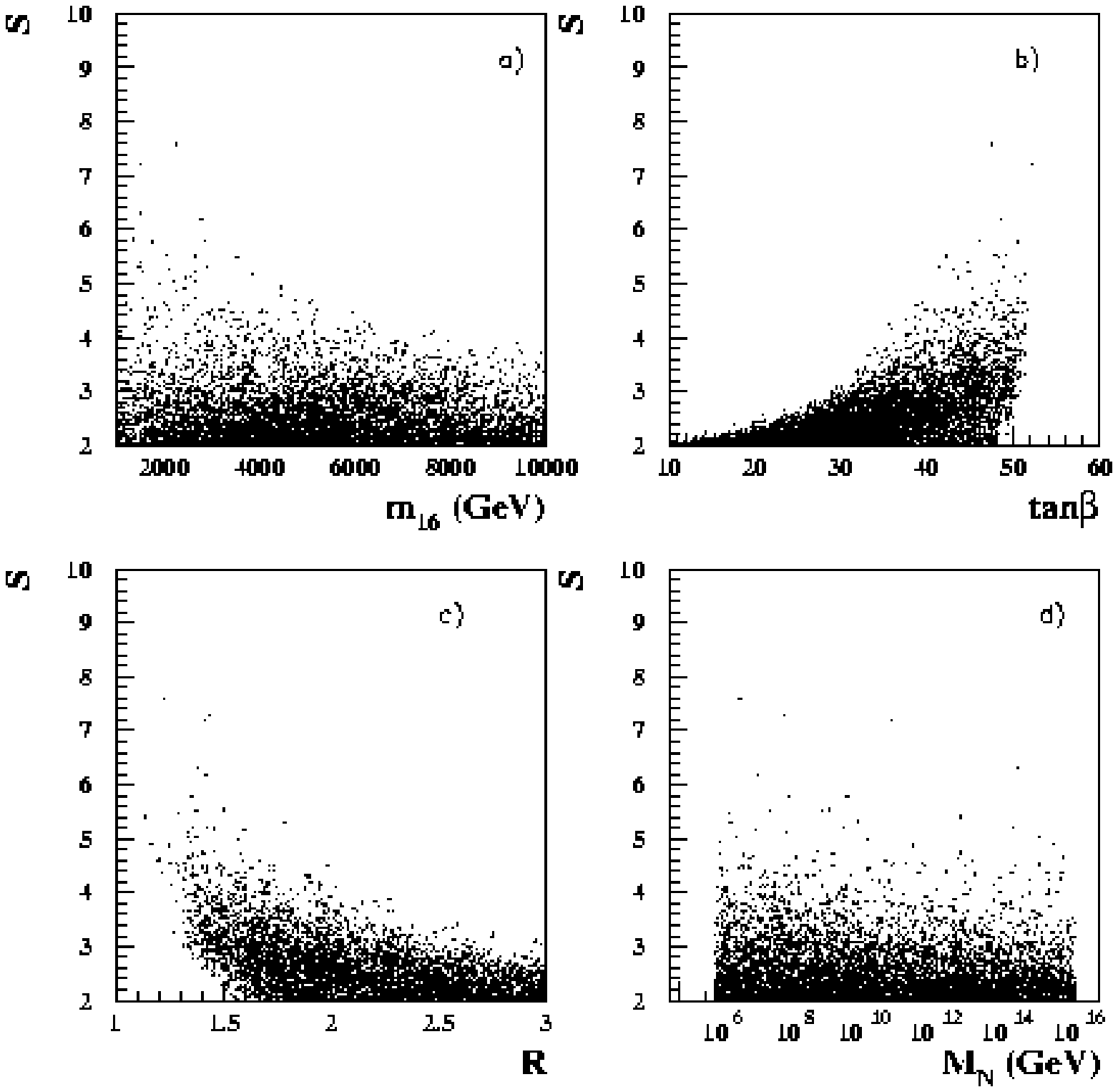}
\caption[]{
A plot of crunch factor $S$ versus {\it a}) $m_{16}$, {\it b}) $\tan\beta$,
{\it c}) $R$ and {\it d}) $M_N$, for all models with $A_0<0$ and $\mu >0$.}
\label{imh2_3}
\end{figure}
\begin{figure}
\dofig{5in}{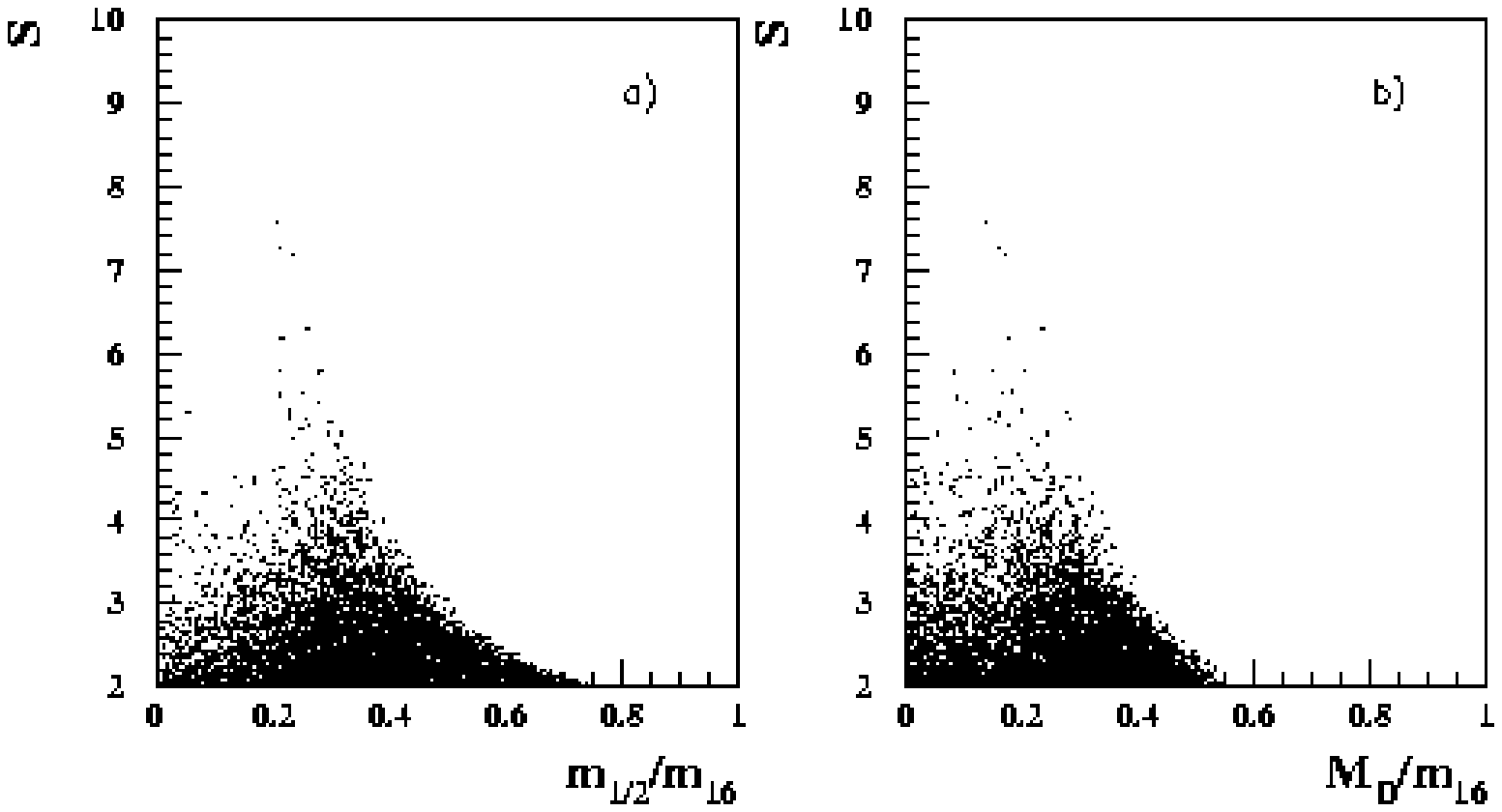}
\caption[]{
A plot of crunch factor $S$ versus {\it a}) $m_{1/2}/m_{16}$ and
{\it b}) $M_D/m_{16}$, for all models with $A_0<0$ and $\mu >0$.}
\label{imh2_4}
\end{figure}
\begin{figure}
\dofig{5in}{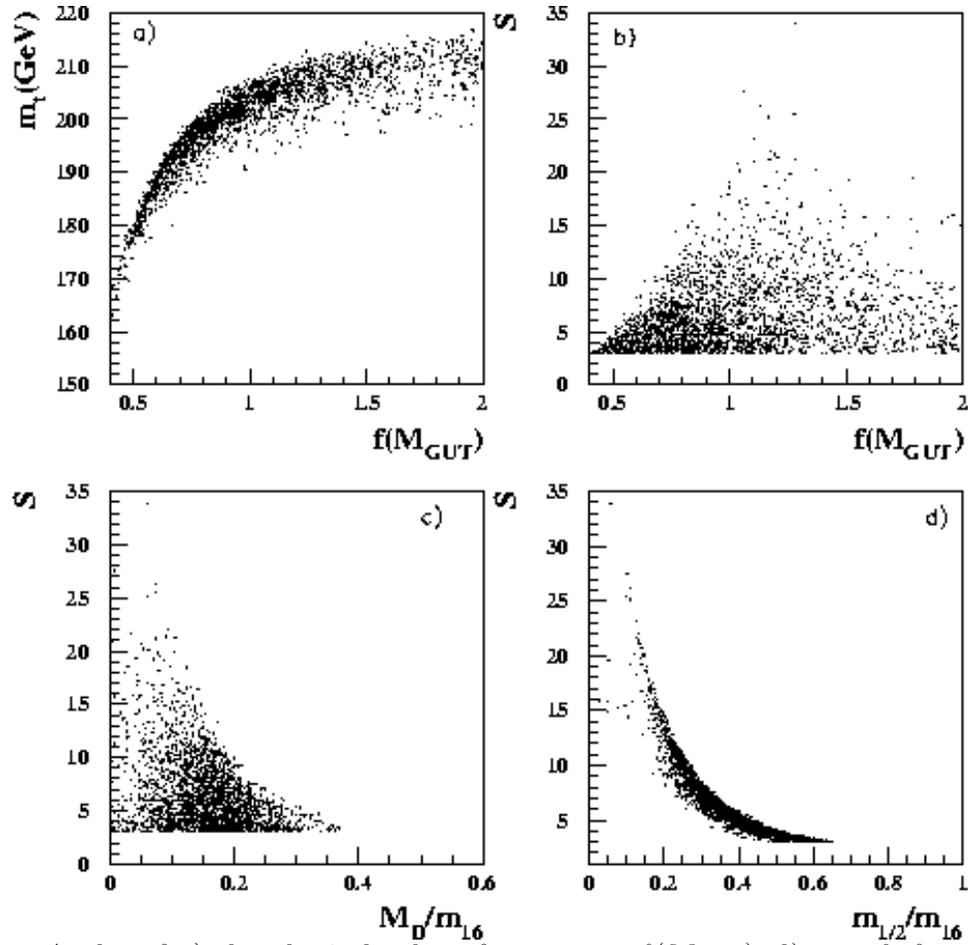}
\caption[]{
A plot of {\it a}) the physical value of $m_t$ versus $f(M_{GUT})$,
{\it b}) crunch factor $S$ versus $f(M_{GUT})$, and
{\it c}) $S$ versus $M_D/m_{16}$ and {\it d}) $S$ versus $m_{1/2}/m_{16}$,
using a top-down solution to RGEs, and a unified $GUT$ scale Yukawa coupling.
In these plots, we adopt $\mu <0$.}
\label{tdmum}
\end{figure}
\begin{figure}
\dofig{5in}{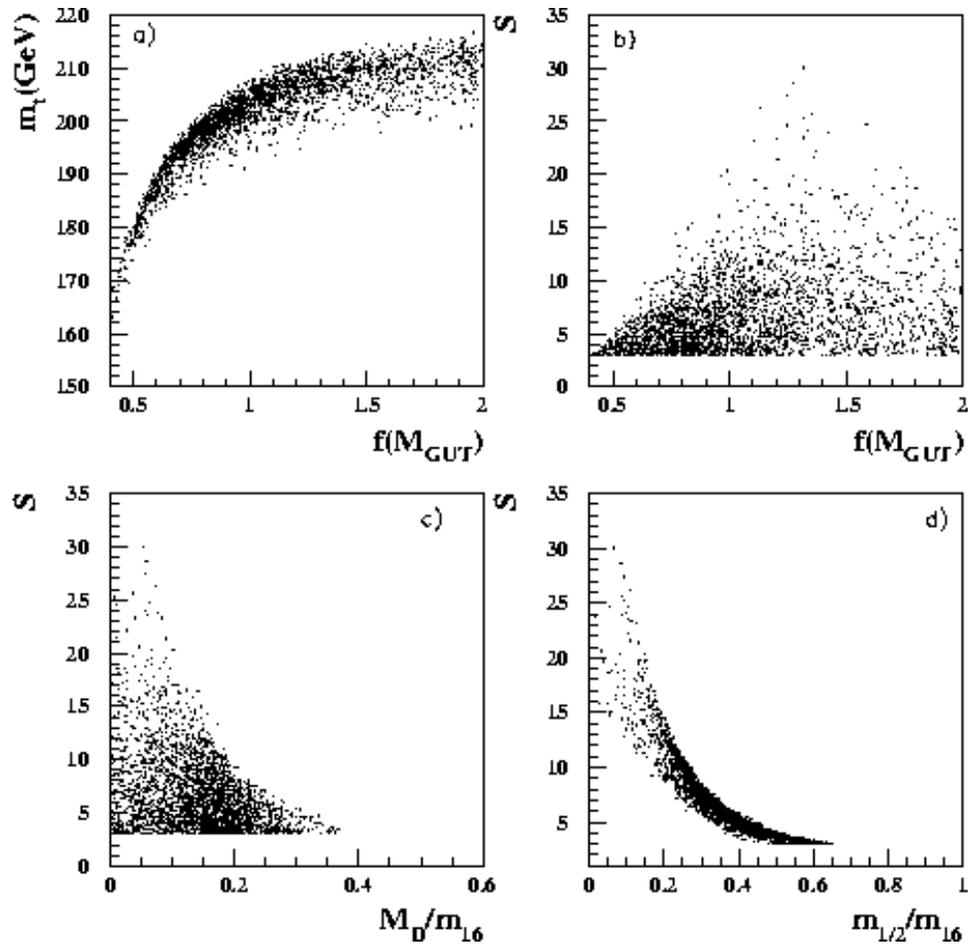}
\caption[]{
Same as Fig. \ref{tdmum}, except for $\mu >0$.}
\label{tdmup}
\end{figure}
\begin{figure}
\dofig{5in}{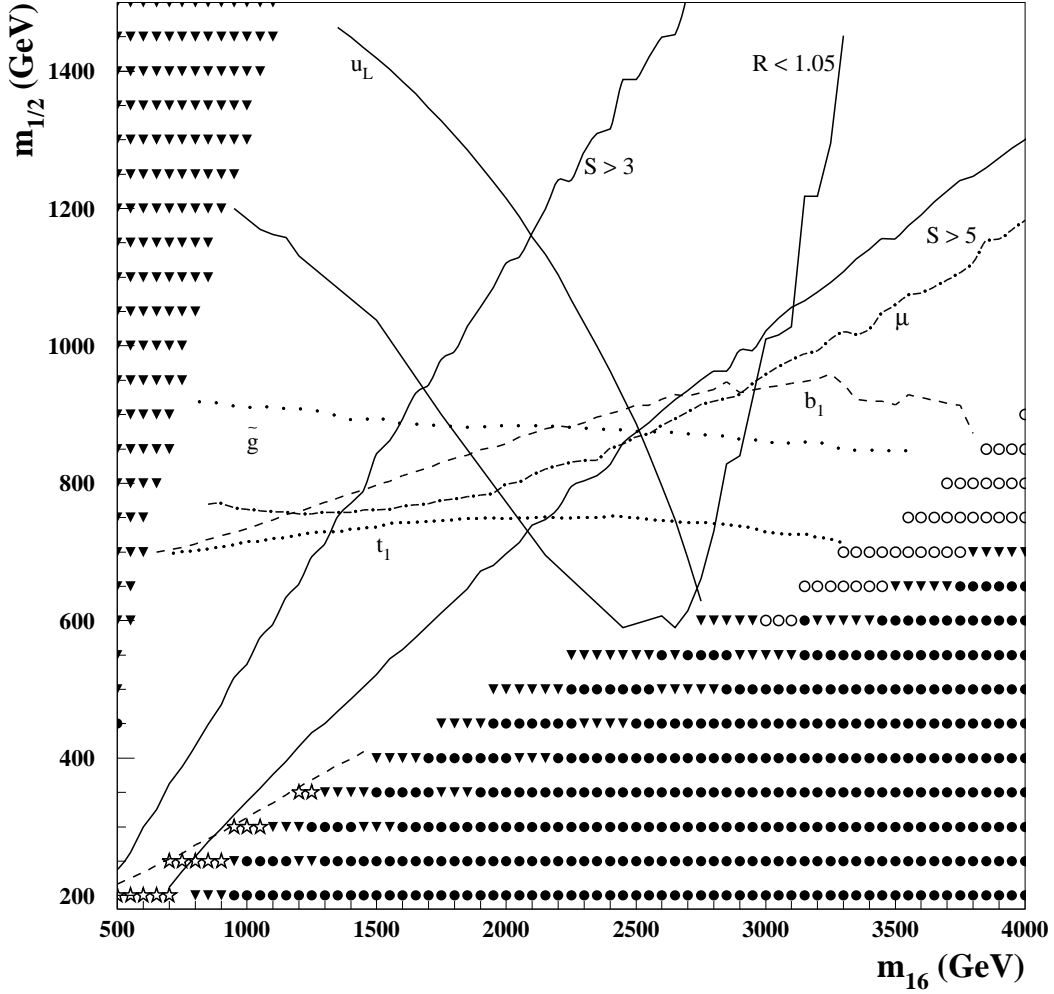}
\caption[]{
A plot of allowed regions for the RIMH model in the
$m_{16}\ vs. m_{1/2}$ plane, for $M_N=1\times 10^7$ GeV, $M_D=0.2 m_{16}$,
$\tan\beta =50$ and $\mu <0$.
Points with triangles are excluded by a non-neutralino LSP, open circles
by lack of REWSB and dots by tachyonic scalar masses.
We show contours of $S=3$ and 5, $R=1.05$, $m_{\tg}=2$ TeV, $m_{\tu_L}=3$ TeV,
$\mu =-1$ TeV, and $m_{\tst_1}$ and $m_{\tb_1}$ of 1 TeV. The stars
denote points where bottom squarks should be detectable at the Fermilab
Tevatron, assuming an integrated luminosity of 25~$fb^{-1}$.}
\label{imh2_5}
\end{figure}
\begin{figure}
\dofig{5in}{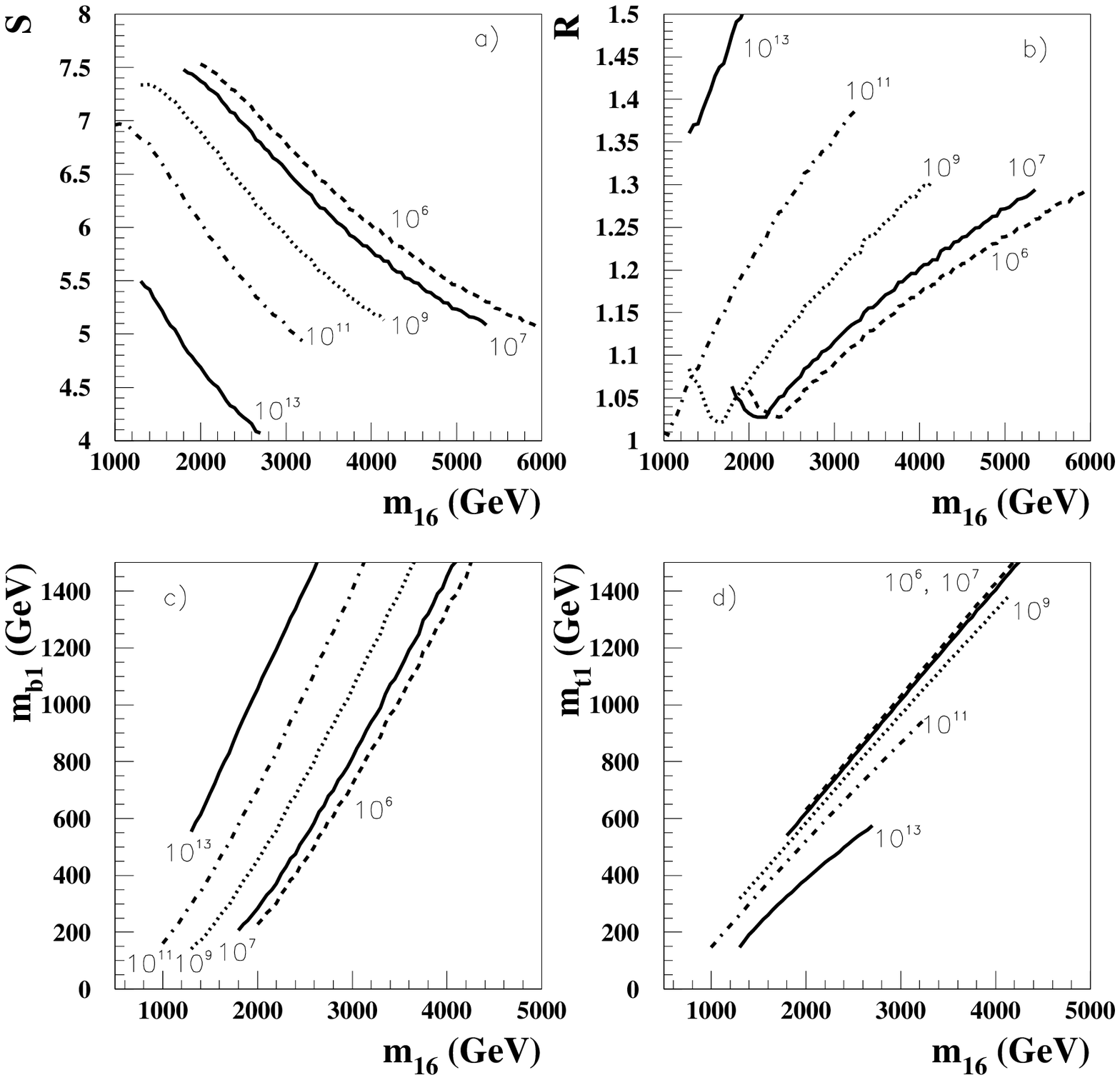}
\caption[]{
A plot of {\it a}) $S$, {\it b}) $R$, {\it c}) $m_{\tb_1}$ and
{\it d}) $m_{\tst_1}$ versus
$m_{16}$ for $m_{1/2}=0.25 m_{16}$, $\tan\beta =50$ and $\mu <0$,
for the various $M_N$ values (in GeV) exhibited in the figure.}
\label{imh2_6}
\end{figure}
\begin{figure}
\dofig{5in}{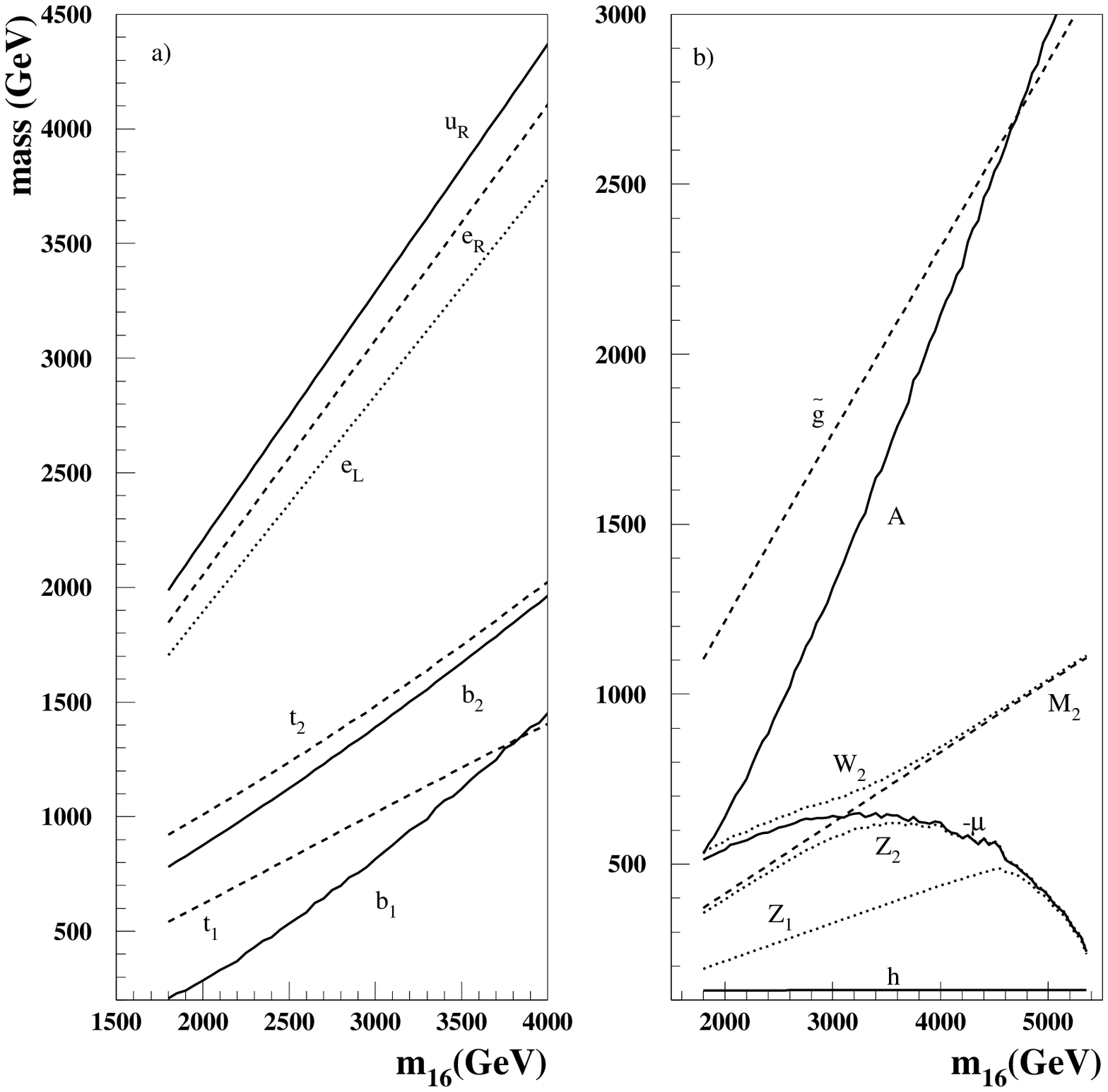}
\caption[]{
A plot of various sparticle and Higgs boson masses and $-\mu$ versus
$m_{16}$ for $m_{1/2}=0.25 m_{16}$, $\tan\beta =50$, $\mu <0$ and
$M_N=1\times 10^7$ GeV.}
\label{imh2_mass1}
\end{figure}
\begin{figure}
\dofig{5in}{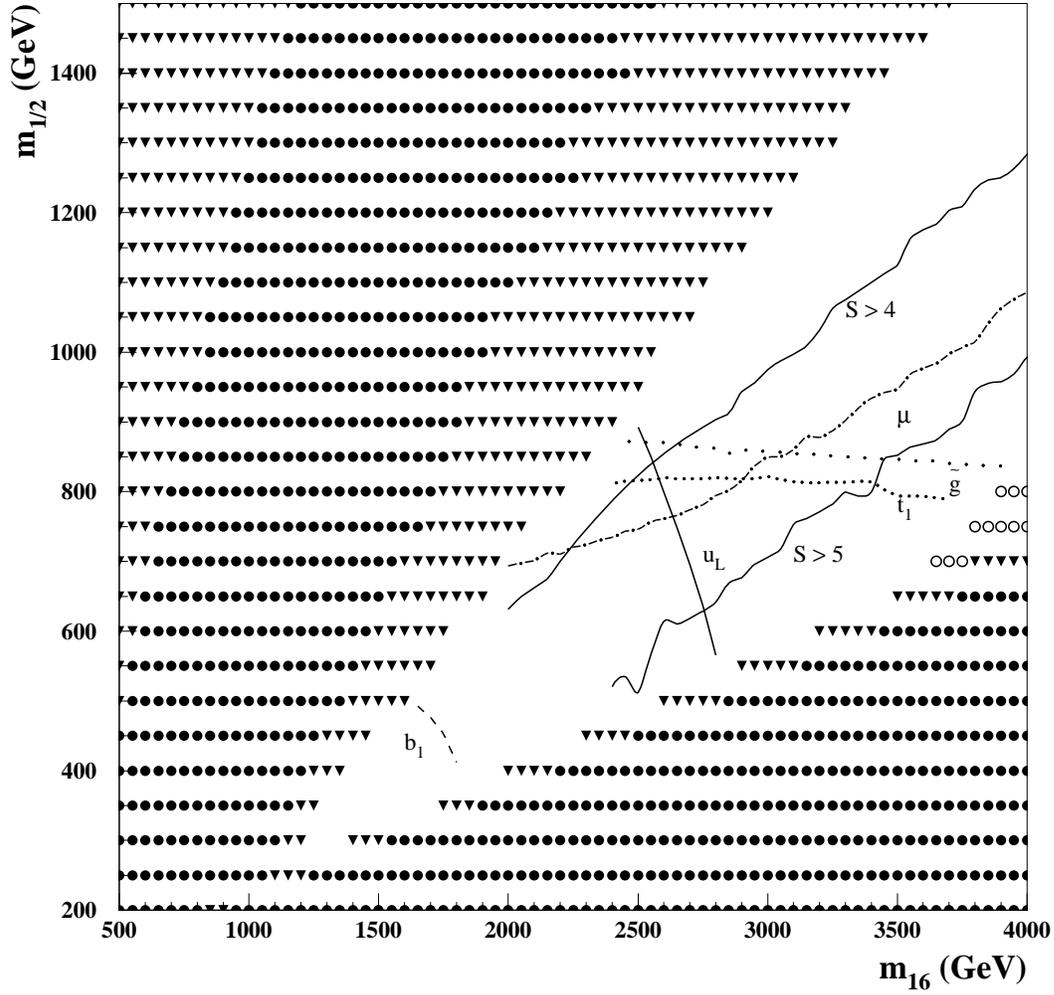}
\caption[]{
Same as Fig. \ref{imh2_5}, except $\mu >0$. The degree of Yukawa unification
ranges from $R=1.6-1.9$ thoughout the plane.}
\label{imh2_7}
\end{figure}
\begin{figure}
\dofig{5in}{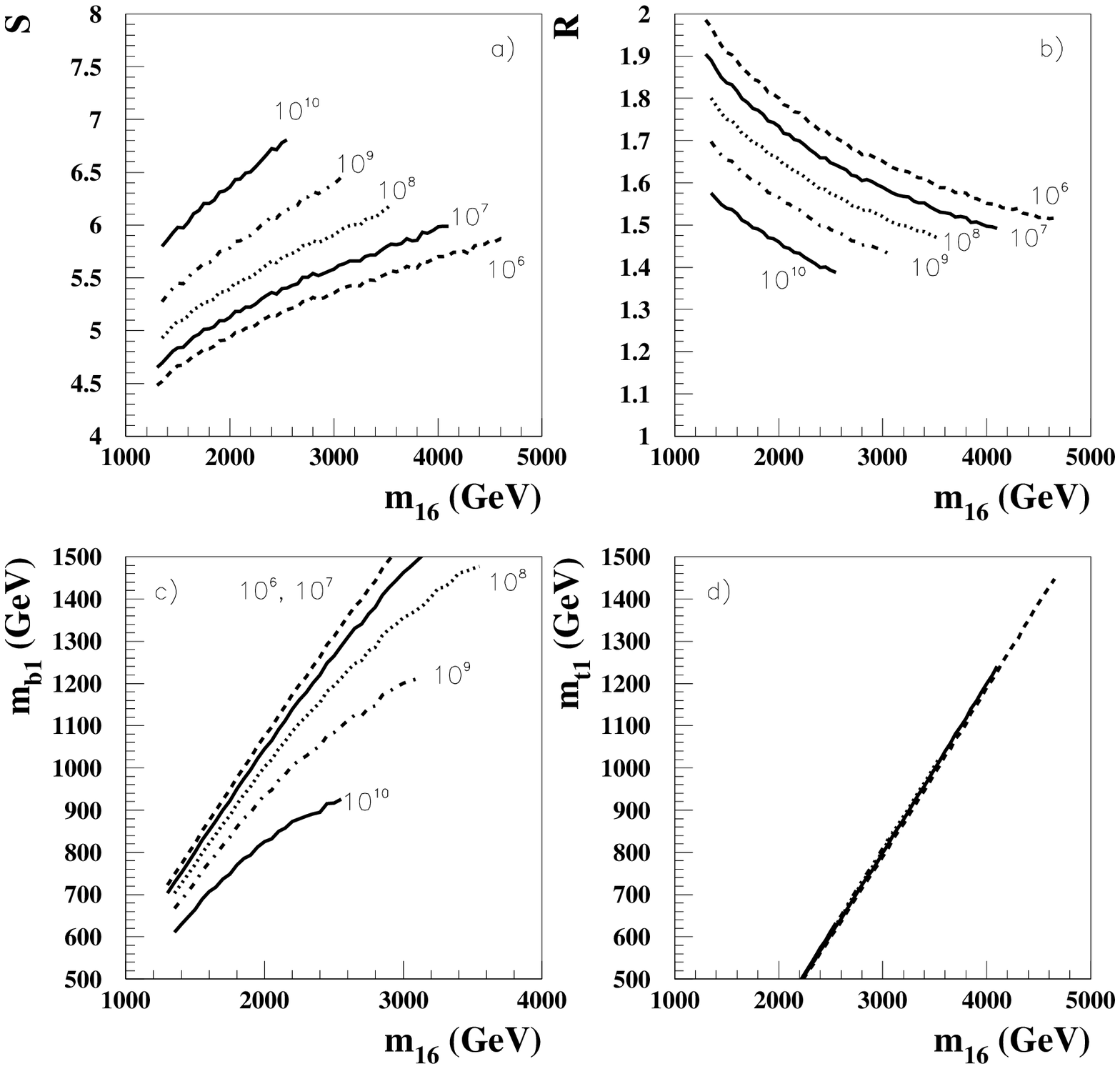}
\caption[]{
A plot of {\it a}) $S$, {\it b}) $R$, {\it c}) $m_{\tb_1}$ and
{\it d}) $m_{\tst_1}$ versus
$m_{16}$ for $m_{1/2}=0.22 m_{16}$, $\tan\beta =50$ and $\mu >0$,
for the various $M_N$ values (in GeV) exhibited in the figure.}
\label{imh2_8}
\end{figure}
\begin{figure}
\dofig{5in}{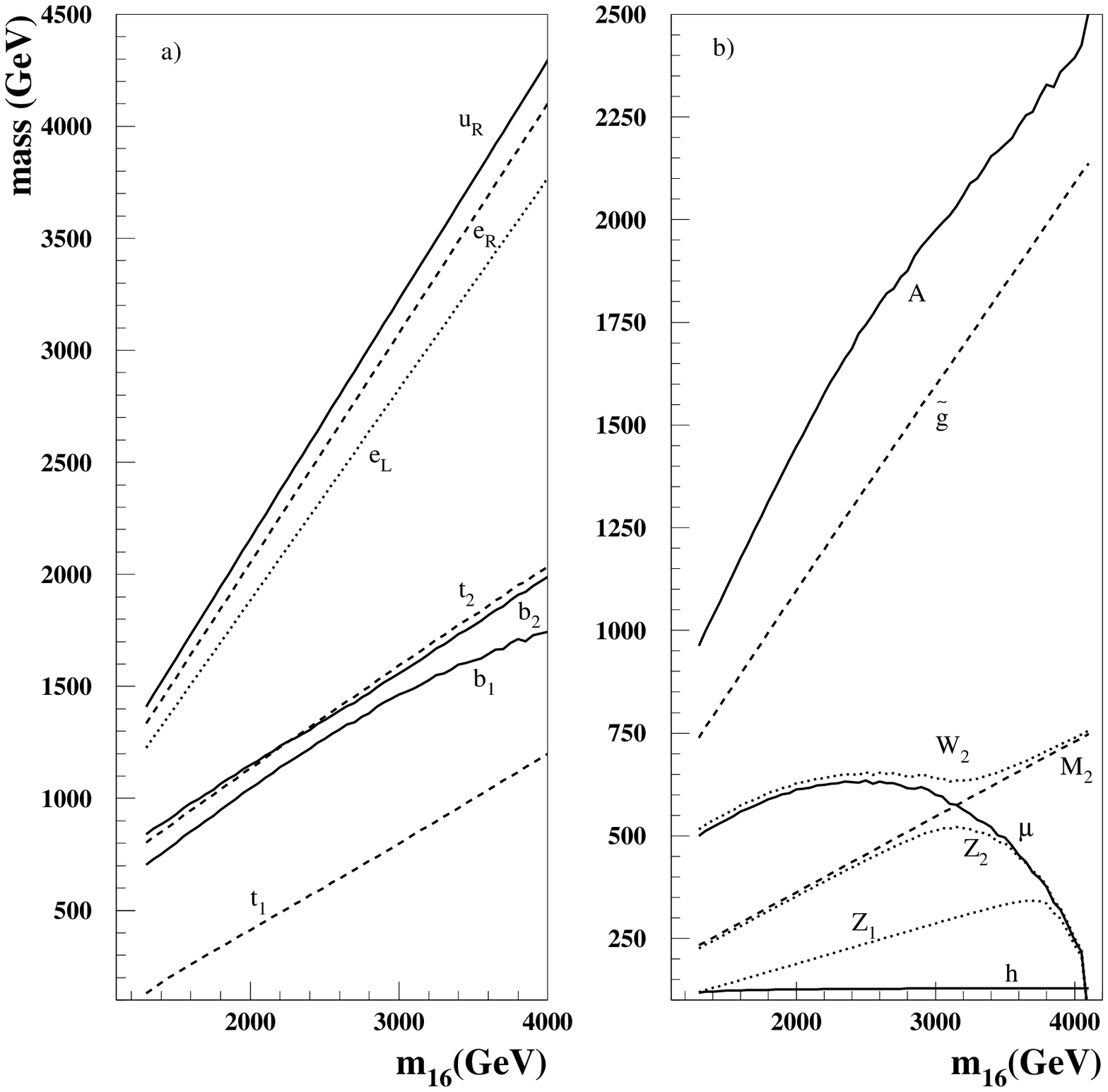}
\caption[]{
A plot of various sparticle and Higgs boson masses and $\mu$ versus
$m_{16}$ for $m_{1/2}=0.22 m_{16}$, $\tan\beta =50$, $\mu >0$ and
$M_N=1\times 10^7$ GeV.}
\label{imh2_mass2}
\end{figure}
\begin{figure}
\dofig{5in}{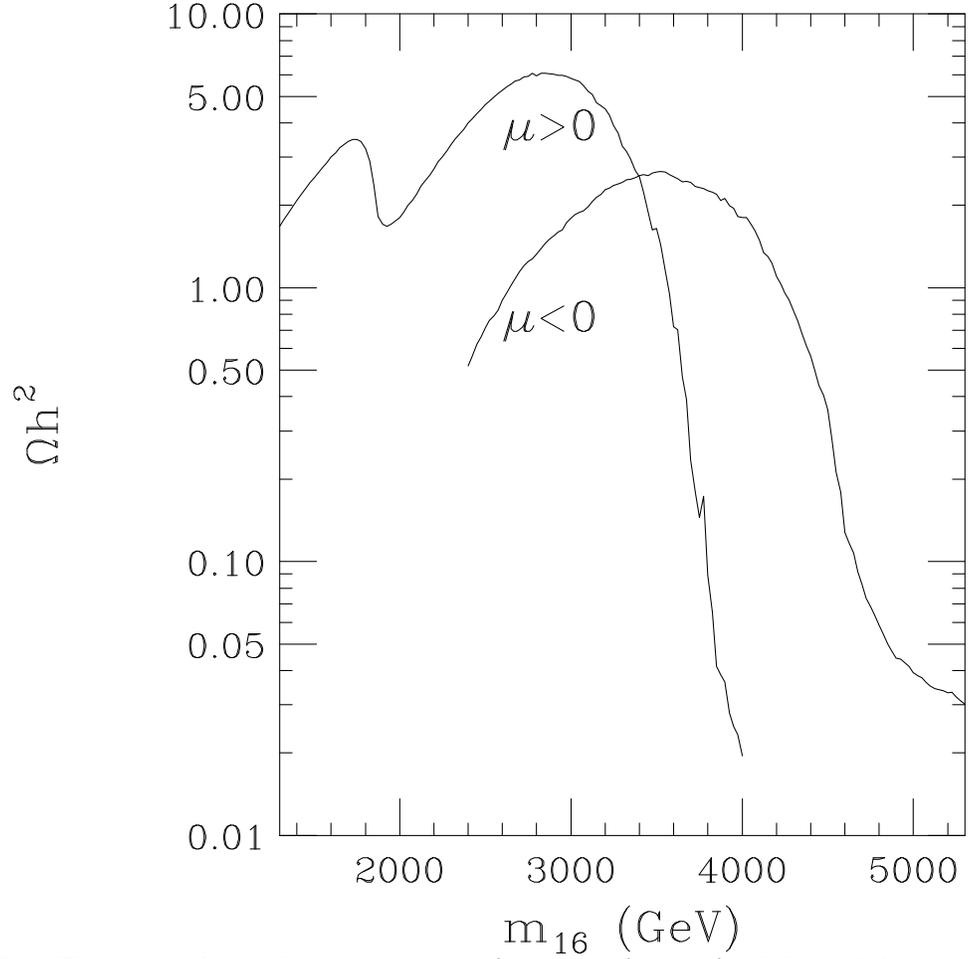}
\caption[]{
The neutralino relic density as a function of $m_{16}$, for
$M_D=0.2 m_{16}$, $\tan\beta =50$, $M_N=1\times 10^7$ GeV
and {\it a}) $\mu <0$ with $m_{1/2}=0.25 m_{16}$ and {\it b}) $\mu >0$
with $m_{1/2}=0.22 m_{16}$.}
\label{imh2_rd}
\end{figure}
\begin{figure}
\dofig{5in}{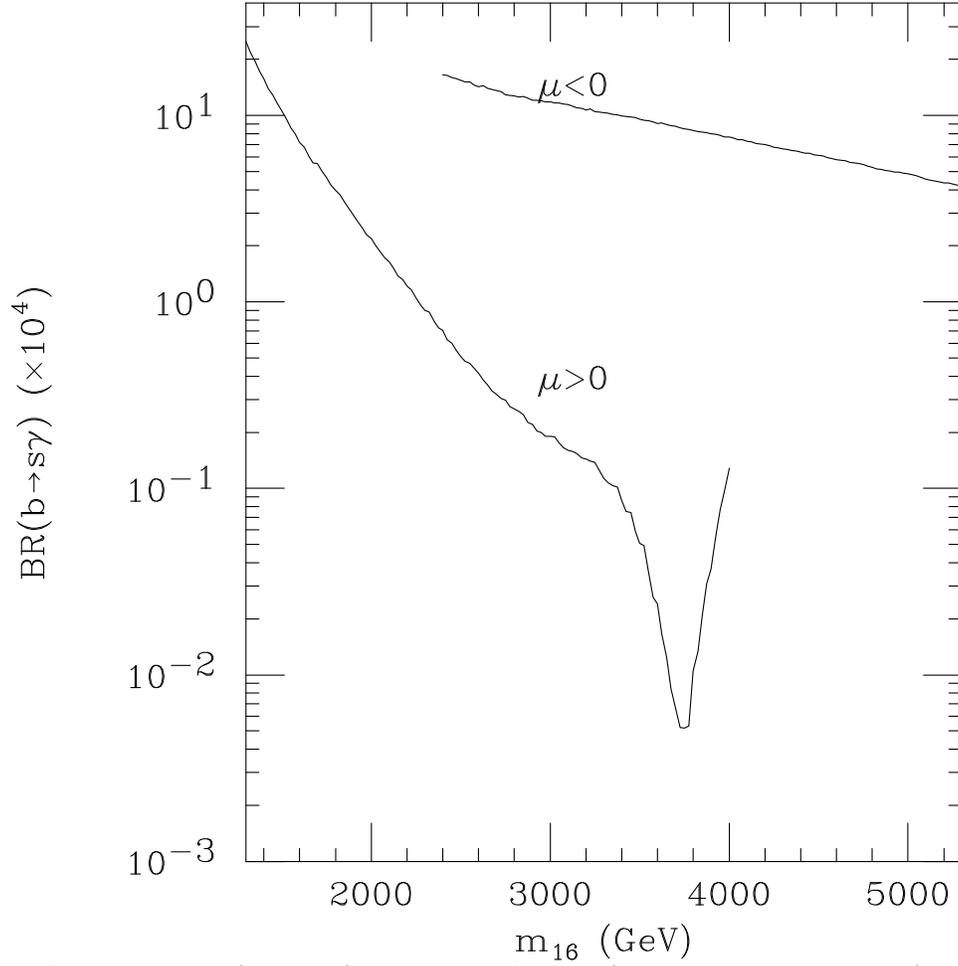}
\caption[]{
The branching fraction for the decay $b\to s\gamma$ for the same two
slices of the parameter plane as in Fig.~\ref{imh2_rd} versus the
parameter $m_{16}$. The CLEO experiment requires that this branching
fraction is between $2\times 10^{-4}$ and $4.5\times 10^{-4}$ at the
95\% CL, while the recent measurement from the BELLE collaboration has a
central value only slightly higher.
}
\label{figbsg}
\end{figure}
\begin{figure}
\dofig{5in}{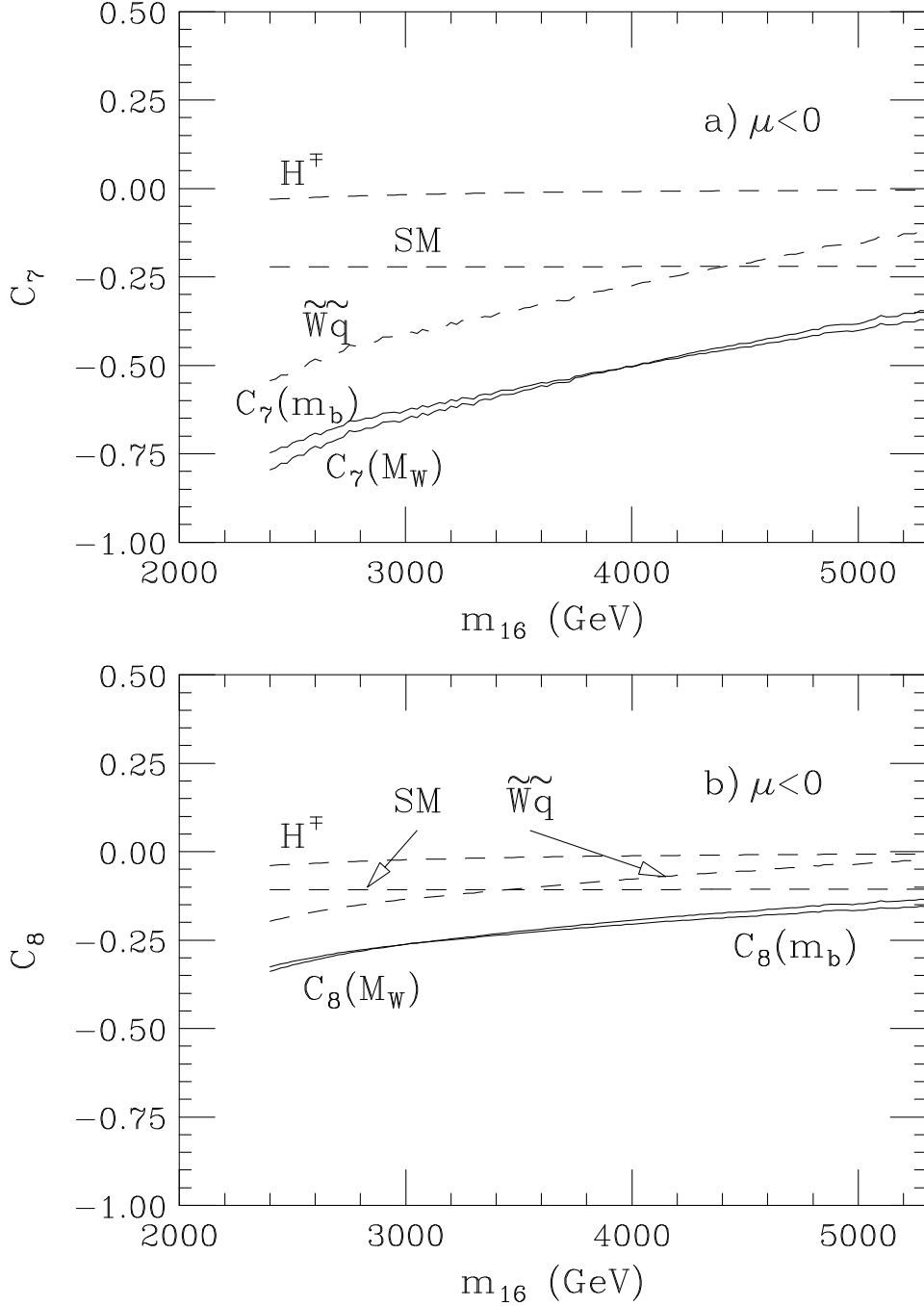}
\caption[]{The Wilson coefficients {\it a})~$C_7$, and {\it b})~$C_8$
that enter the computation of the radiative decay of the $b$ quark,
versus the model parameter $m_{16}$ for the parameter space slice with
negative $\mu$. The dashed lines labelled, SM, $H^{\mp}$ and $\tw$,
respectively denote contributions to $C_7$ or $C_8$ from the SM $tW$
loop, the $tH^+$ loop and the $\tw_i\tq_j$ loop. The sum of these is
shown by the solid line labelled $C_{7,8}(M_W)$. The result of evolving
this down to the scale $m_b$ is shown as the other solid line labelled
$C_{7,8}(m_b)$.
}
\label{bsgn}
\end{figure}
\begin{figure}
\dofig{5in}{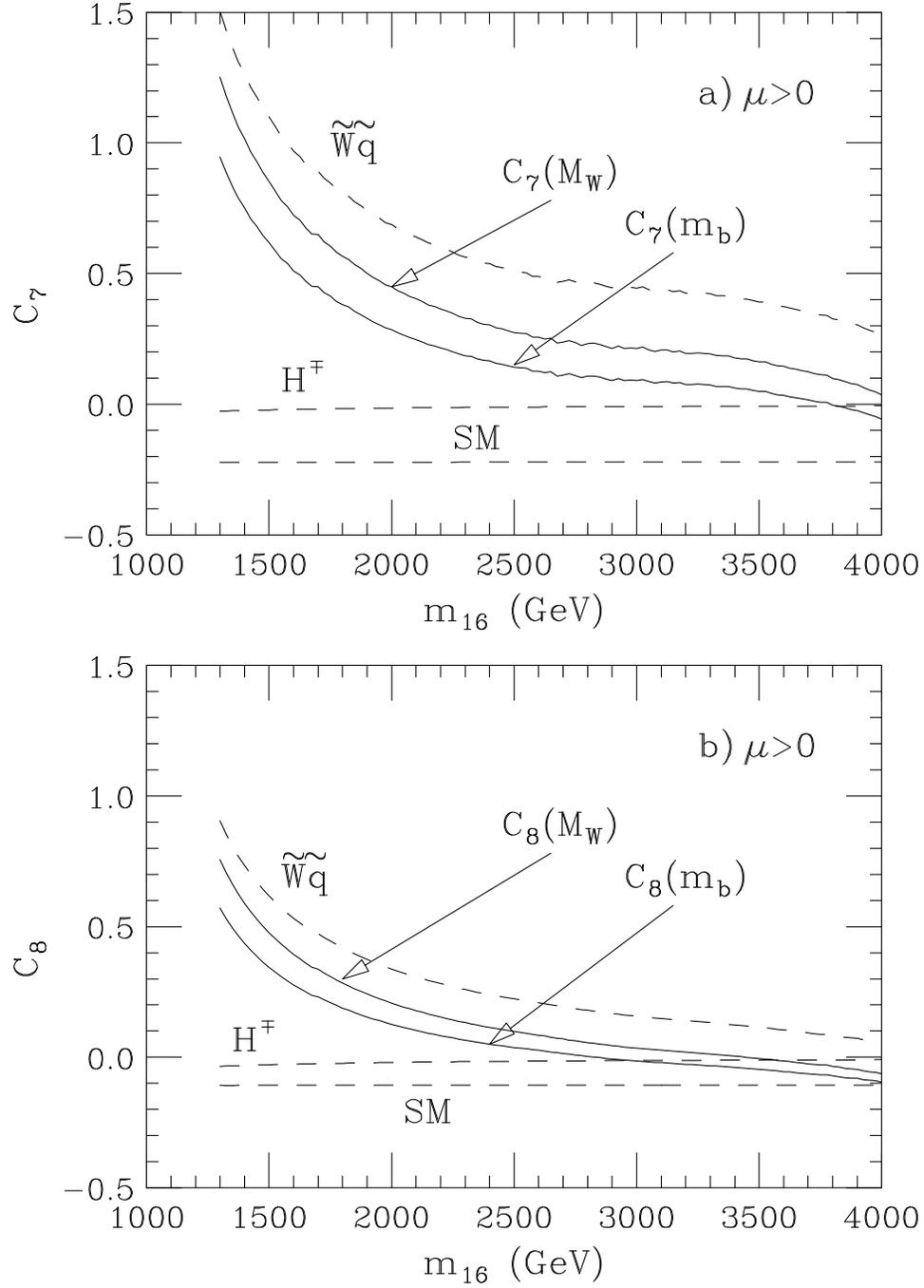}
\caption[]{The same as Fig.~\ref{bsgn}, except for the parameter space
slice with $\mu > 0$.
}
\label{bsgp}
\end{figure}
\begin{figure}
\dofig{6in}{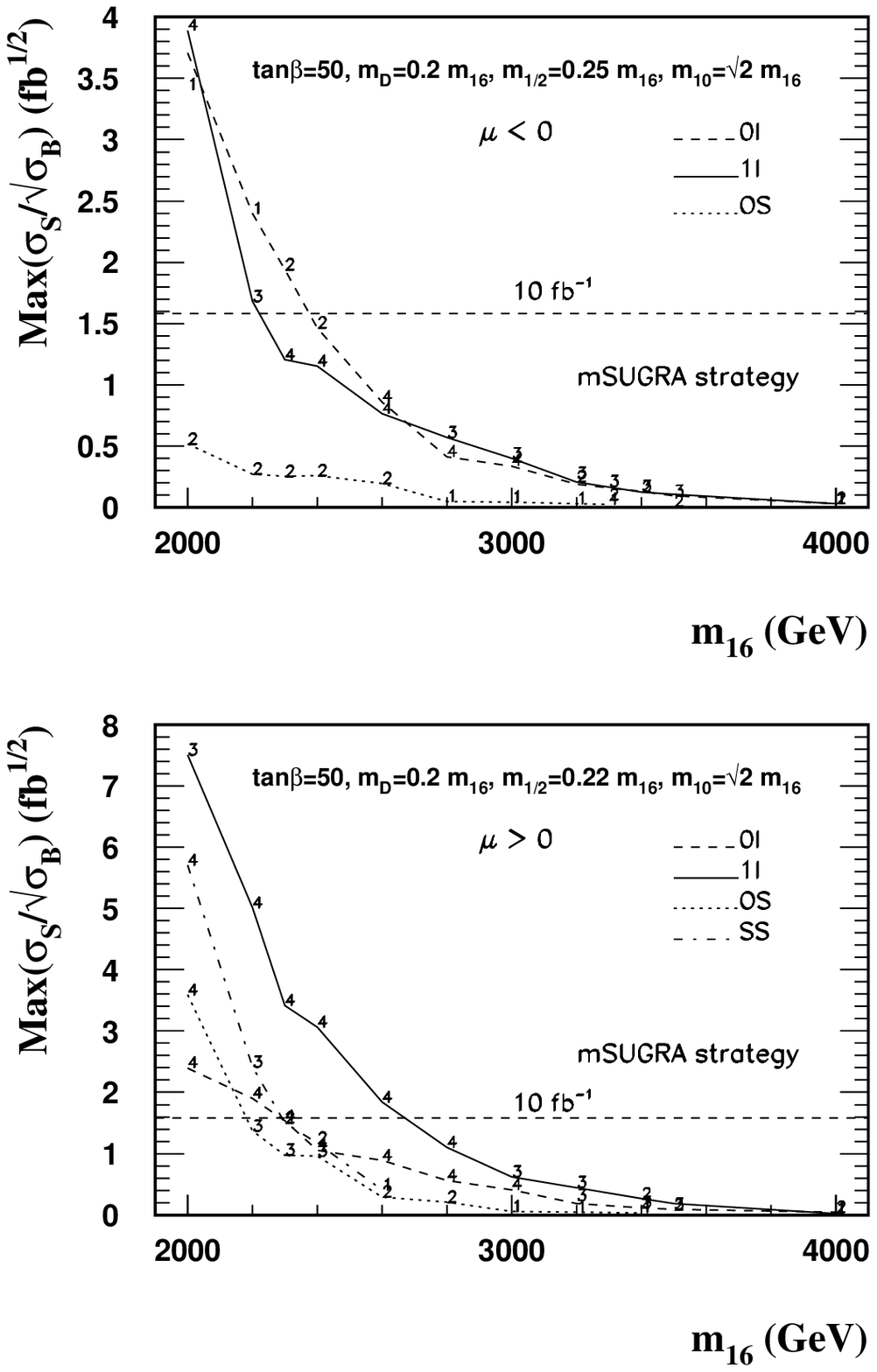}
\caption[]{
A plot of signal cross section over square root of background
cross section for $0\ell$, $1\ell$, $OS$ and $SS$ dilepton events after cuts
from the RIMH model at the CERN LHC, versus the parameter $m_{16}$.
The $n-\sigma$ significance can be gained by multiplying by the square root of
the integrated luminosity in fb.
In {\it a}), we take $m_{1/2}=0.25 m_{16}$ and $\mu <0$.
In {\it b}), we take $m_{1/2}=0.22 m_{16}$ and $\mu >0$.
In both frames, we take $M_D=0.2 m_{16}$, $\tan\beta =50$ and
$M_N=1\times 10^7$ GeV.}
\label{yili1}
\end{figure}
\begin{figure}
\dofig{5in}{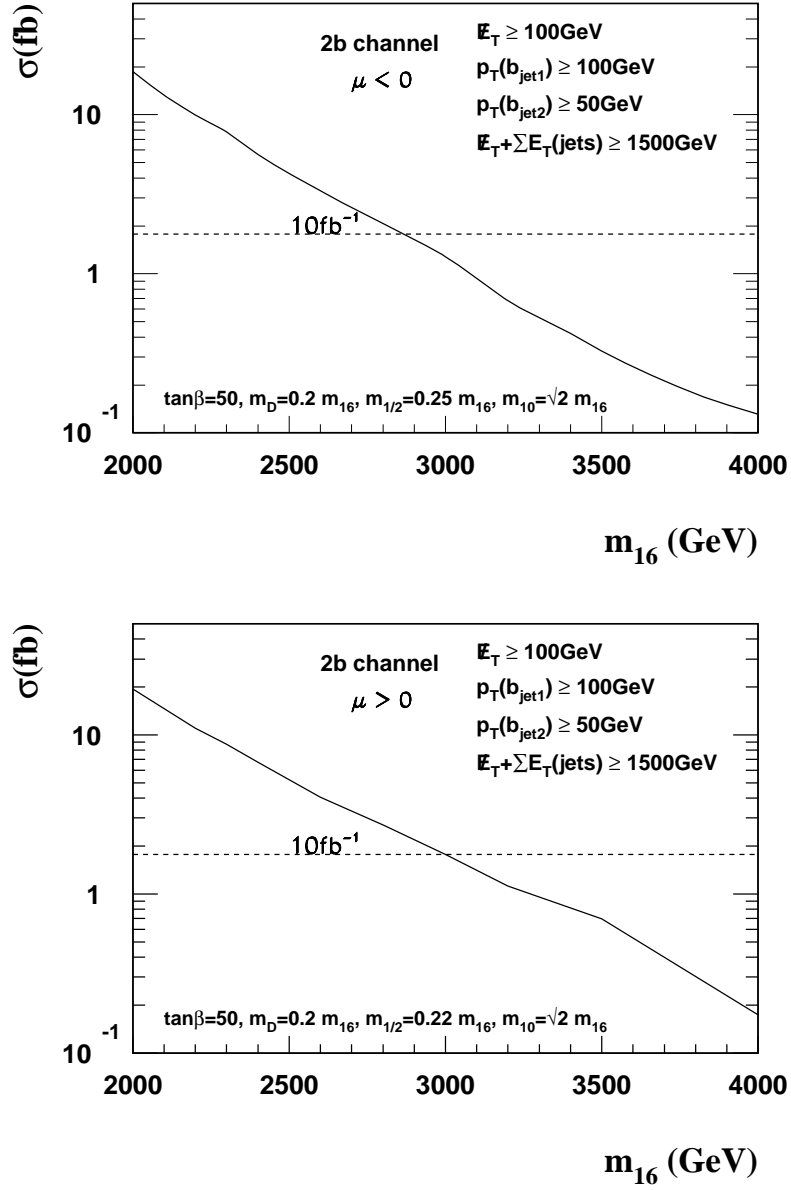}
\caption[]{
A plot of signal cross section for multijet $+\eslt$ events
which contain exactly two tagged $b$-jets, after cuts,
from the RIMH model at the CERN LHC, versus the parameter $m_{16}$.
In {\it a}), we take $m_{1/2}=0.25 m_{16}$ and $\mu <0$.
In {\it b}), we take $m_{1/2}=0.22 m_{16}$ and $\mu >0$.
In both frames, we take $M_D=0.2 m_{16}$, $\tan\beta =50$ and
$M_N=1\times 10^7$ GeV.}
\label{yili2}
\end{figure}

\end{document}